\documentclass[12pt]{article}
\usepackage{apalike}\usepackage{url}\usepackage{hyperref}\usepackage{cite}
\hypersetup{
  colorlinks=true,
  linkcolor=blue,
  filecolor=magenta,      
  urlcolor=cyan,
  pdftitle={Overleaf Example},
  pdfpagemode=FullScreen,
}
\usepackage[utf8]{inputenc}\usepackage{enumerate}\usepackage{amsmath}\usepackage{amssymb}\usepackage{amsthm}\usepackage{graphicx}\usepackage[dvipsnames]{xcolor}\usepackage{tikz}\usepackage{bm}\usepackage{proof}\usepackage{amsfonts}\usepackage{tcolorbox}\usepackage{ascmac}\usepackage{fancybox}
\usepackage[all]{xy}\usepackage{varwidth}\usepackage{etoolbox}\usepackage{siunitx}\usepackage{physics}
\AtBeginDocument{\RenewCommandCopy\qty\SI}  %% told to add this by compiler 7/27/2023
\hypersetup{
    colorlinks,
    citecolor=black,
    filecolor=black,
    linkcolor=black,
    urlcolor=black
}
\usepackage[edges]{forest}
\usepackage{tree}
\newtheorem{definition}{Definition}[section]
\newtheorem{theorem}{Theorem}[section]
\newtheorem{proposition}{Proposition}[section]
\newtheorem{corollary}{Corollary}[section]

\newtheorem{assumption}{Assumption}[section]

\theoremstyle{definition}
\AtEndEnvironment{remark}{\qed}
\newtheorem{remark}{Remark}[section]

\AtBeginDocument{\RenewCommandCopy\qty\SI}

\renewcommand{\P}{\mathbb{P}}
\newcommand{\al}{\alpha}

\usepackage[margin=1in, includefoot, footskip=0.5in]{geometry}
\usepackage{bbm}

\usepackage{natbib}
\usepackage{comment}
\excludecomment{en-text}

\newcommand{\indep}{\perp \!\!\! \perp}
\newcommand{\att}{{\theta_{\text{ATT}}}}
\newcommand{\naive}{{\theta^{\text{na\"ive}}}}
\newcommand{\atte}{{\theta_{\text{ATT}}^{\text{E}}}}

\newcommand{\attlu}{{\theta_{\text{ATT}}^{\text{LU}}}}
\newcommand{\attec}{{\theta_{\text{ATT}}^{\text{ECB}}}}
\newcommand{\atee}{{\theta_{\text{ATE}}^{\text{E}}}}
\newcommand{\ateo}{{\theta_{\text{ATE}}^{\text{O}}}}
\newcommand{\atelu}{{\theta_{\text{ATE}}^{\text{LU}}}}
\newcommand{\ateec}{{\theta_{\text{ATE}}^{\text{ECB}}}}
\newcommand{\E}{{\text{E}}}
\renewcommand{\P}{{\text{P}}}

\begin{document}
\allowdisplaybreaks

\title{\Large A Bracketing Relationship for Long-Term Policy Evaluation\\ with Combined Experimental and Observational Data\thanks{\setlength{\baselineskip}{4mm}We would like to thank Raj Chetty for generously allowing us to use his data sets for our empirical application. We benefitted from useful comments by  Raj Chetty, Jamie Fogel, Hidehiko Ichimura, Matt Staiger, and Jimmy Stratton. All remaining errors are ours.\smallskip}}
\author{
Yechan Park\thanks{\setlength{\baselineskip}{4mm}Opportunity Insights, Harvard University, 1280 Massachusetts Avenue, Cambridge, MA 02138 Email: \texttt{yechanpark@fas.harvard.edu}\smallskip}
\and
Yuya Sasaki\thanks{\setlength{\baselineskip}{4mm}Brian and Charlotte Grove Chair and Professor of Economics. Department of Economics, Vanderbilt University, VU Station B \#351819, 2301 Vanderbilt Place, Nashville, TN 37235-1819 Email: \texttt{yuya.sasaki@vanderbilt.edu}}
}
\date{}

\maketitle
\begin{abstract}\setlength{\baselineskip}{6mm}
 Combining short-term experimental data with observational data enables credible long-term policy evaluation. The literature offers two key but non-nested assumptions, namely the latent unconfoundedness \citep[LU;][]{athey2020combining} and equi-confounding bias \citep[ECB;][]{ghassami2022combining} conditions, to correct observational selection. Committing to the wrong assumption leads to biased estimation.
To mitigate such risks, we provide a novel \textit{bracketing relationship} \citep[cf.][]{angrist2009mostly} repurposed for the setting with data combination: the LU-based estimand and the ECB-based estimand serve as the lower and upper bounds, respectively, with the true causal effect lying in between if \textit{either} assumption holds. For researchers further seeking point estimates, our Lalonde-style exercise suggests the conservatively more robust LU-based lower bounds align closely with the hold-out experimental estimates for educational policy evaluation. We investigate the economic substantives of these findings through the lens of a nonparametric class of selection mechanisms and sensitivity analysis. We uncover as key the sub-martingale property and sufficient-statistics role \citep{chetty2009sufficient} of the \textit{potential outcomes} of student test scores \citep{chetty2011does,chetty2014measuring1}.
%{(165 words)}
    
\begin{comment}
The literature on long-term policy evaluation with data combination proposes alternative identifying assumptions, namely the latent unconfoundedness (LU) and equi-confounding bias (ECB).
They are mutually non-nested and lead to distinct identifying formulae.
In this light, we introduce a novel \textit{bracketing relationship} \citep[cf.][]{angrist2009mostly} repurposed for the setting with data combination: the LU-based estimand is less than or equal to the ECB-based estimand.
Thus, if the LU condition \textit{or} the ECB condition holds, then the true treatment effect is bounded below by the LU-based estimand and bounded above by the ECB-based estimand.
Empirical applications using Project Star and extensive administrative records validate the bracketing relation and reveal their informativeness on the sign and magnitude of the causal effects. A LaLonde-style analysis suggests that the LU-based estimates more accurately reflect experimental estimates in the context of educational policy evaluation.
{(145 words)}
\end{comment}


%\medskip\noindent
%{\bf JEL Codes:}
\end{abstract}

\newpage
%%%%%%%%%%%%%%%%%%%%%%%%%%%%%%%%%%%%%%%
\section{Introduction}
%%%%%%%%%%%%%%%%%%%%%%%%%%%%%%%%%%%%%%%

In devising, implementing, and evaluating a policy, it is essential to precisely estimate its long-term impacts. 
Experimental data free of confounding have been successful to this end \citep[e.g.,][]{card1992does,hotz2005predicting}. 
However, collecting long-term experimental data is costly and time-consuming. 

There has been work to overcome this challenge by combining short-term experimental data and long-term observational data.
This literature has heavily relied on the so-called \textit{surrogacy condition} \citep{prentice1989surrogate}, requiring the long-term outcome to be conditionally independent of the treatment given a surrogate.
However, subsequent research has questioned the validity of this assumption, leading to discussions referred to as the \textit{surrogate paradox} %\citep{freedman1992statistical,buyse1998criteria,buyse2000validation,frangakis2002principal,chen2007criteria} 
-- see \cite{vanderweele2013surrogate} for a comprehensive exposition.
While several approaches have been proposed to mitigate the surrogacy problem \citep[e.g.,][]{athey2019surrogate,ma2021individual}, a particularly promising breakthrough was made by \citet{athey2020combining} with their novel \textit{latent unconfoundedness} (LU) condition, replacing the conventional surrogacy condition, to correct for endogenous treatment selection in observational data.

Together with the standard internal and external validity assumptions for experiments, the LU condition proposed by \citet{athey2020combining} allows one to identify long-term treatment effects with combined experimental and observational data.
More recently, \citet{ghassami2022combining} propose an alternative assumption to correct for selection, called the \textit{equi-confounding bias} (ECB) condition.\footnote{\label{foot:additional}\citet{ghassami2022combining} propose several alternative approaches including the ECB. The others (e.g., Bespoke instrumental variables and proximal data fusion) require additional information besides the baseline setup, and hence we will not compare them in this paper. See also \citet{imbens2022long} who use additional data (multiple short-term outcomes with a sequential structure) to invoke new assumptions that tackle persistent confounding.}
These two alternative assumptions are non-nested and lead to distinct identifying formulae in general.
As such, assuming the ECB condition leads to biased estimates when the LU condition is true, and vice versa.

In this light, we investigate a \textit{bracketing relationship} \citep[cf.][Sections 5.3--5.4]{angrist2009mostly} between the two distinct identifying formulae implied by the LU and ECB conditions.
Namely, we show that the LU-based estimand is less than or equal to the ECB-based estimand. 
Thus, if \textit{either} the LU condition \textit{or} the ECB condition is true, then we enjoy doubly-robust bounds for the true long-term treatment effect, bounded below by the LU-based estimand and bounded above by the ECB-based estimand.
This result implies that both the LU and ECB approaches are useful, and should not be treated as mutually exclusive alternatives in practice.
If a researcher seeks a point estimand and puts a higher priority on Type I Error than Type II Error, then the LU-based estimand is conservatively more robust in the sense that it underestimates the true long-term treatment effect.

\begin{comment}
weakly underestimates the true non-negative long-term treatment effect when the ECB condition is true.
On the other hand, we show that the ECB-based estimand weakly overestimates it when the LU condition is true.
Since underestimation is a more conservative direction of bias than overestimation, our result implies that it is conservatively more robust to consider the LU-based estimand in the sense that scientists put a higher priority on Type I Error than Type II Error.
\end{comment}

Through an empirical analysis using Project Star and extensive administrative data, we demonstrate that our derived bracketing relation indeed holds. 
Furthermore, an analysis \textit{a la} \citet{lalonde1986evaluating} suggests that the LU-based estimates accurately reflect experimental estimates in educational policy evaluation.
On the other hand, the ECB-based estimates are fairly distant from the experimental estimates.
This suggests the potential success of the LU-based and the potential failure of the ECB-based estimates in our data setting. However, we take a step further to provide qualitative reasoning for these empirical findings, further harnessing the information value of the Lalonde-style exercise. We provide economically interpretable \textit{necessary and sufficient} conditions for the two assumptions through the lens of a nonparametric family of selection mechanisms. These conditions are expressed as general restrictions, rather than being tied to any particular parametric model of selection. We find that the LU condition and the ECB condition are essentially equivalent to an invertibility condition and a martingale condition, respectively.
 We thus connect the success of LU condition to the literature on childhood educational interventions \citep{chetty2014measuring1,chetty2011does} documenting the student's past test score as a sufficient statistic \citep{chetty2009sufficient} for latent unobservables.  Crucially, in our new setting under data combination, it is the \textit{latent potential} test score (as in the LU condition), rather than the \textit{observed} test score (as in the conventional surrogacy condition), that plays the role analogous to a sufficient statistic. Likewise, we connect the failure of the ECB condition to the dubious assumption of student test scores evolving as a martingale process.
Finally, using sensitivity analysis techniques, we embed these economic assumptions into our empirical application. Anchoring the hold-out experimental estimate as the baseline truth, we assess how much the qualitative assumptions need to be violated to recover the experimental estimate in our empirical application. We find that the data-generating process is more consistent with a sub-martingale process as opposed to a martingale process, explaining the failure of the ECB-based estimates to replicate the experiment in our context.

We highlight the significance of establishing the bracketing relation between the LU- and ECB-based estimands. 
First, to our knowledge, the LU and ECB conditions are the sole assumptions to point-identify long-term average treatment effects\footnote{\label{foot:quantile_quantile}There is another condition, called the quantile-quantile ECB condition \citep[][Sec. 4.2]{ghassami2022combining} It is useful for identifying distributional and quantile treatment effects, but it will not recover the average effects unless additional assumptions, such as the rank invariance, are imposed.} under our setting without additional data (e.g., sequential outcomes, instrumental variables, or proxy variables).
Second, there are practical implications, since we anticipate the LU condition, as the pioneering assumption for this data setting, and the ECB condition, due to its reliance on the parallel trend assumptions, to be widely adopted in causal analysis in a combined data setting in the future. 
Third, despite the growing theoretical interest in combining experimental and observational data, empirical applications in economics remain limited. 
Demonstrating how informative the bracketing is in our application, while remaining agnostic on the choice of the estimand, may be a promising step to bridge this gap. 

In the broader literature, the LU condition encompasses the lagged dependent variable (LDV) model and the ECB condition is analogous to the parallel trend (PT) condition in the context of the conventional policy evaluation \textit{without} data combination -- see \citet[][Ch. 5]{angrist2009mostly}.
In that classical context \textit{without} data combination, there is also a well-known \textit{bracketing relation} that the LDV regression weakly underestimates the true non-negative treatment effect under the PT assumption while the fixed-effect (or DID) regression weakly overestimates it under the LDV assumption \citep[cf.][Sections 5.3-5.4]{angrist2009mostly} -- also see \citet{ding2019bracketing} for nonparametric bracketing results.
Our contribution to this literature is to provide an analogous \textit{bracketing relation} for the novel setting of combined experimental and observation data for long-term policy evaluation involving more complicated formulae.

\section{Data Combination for Long-Term Policy Evaluation}\label{sec:setup}
%%%%%%%%%%%%%%%%%%%%%%%%%%%%%%%%%%%%%%%
This section introduces the setup following the literature \citep[e.g.,][among others]{athey2020combining,ghassami2022combining} on long-term policy evaluation with combined experimental and observational data.

%%%%%%%%%%%%%%%%%%%%%%%%%%%%%%%%%%%%%%%
\subsection{The Baseline Setup of Data Combination}\label{sec:baseline}
%%%%%%%%%%%%%%%%%%%%%%%%%%%%%%%%%%%%%%%
A researcher conducts a randomized experiment aimed at assessing the effects of a policy.
For each individual indexed by $i$ in the experiment (indicated by $G_i=E$), the researcher observes pre-treatment covariates $X_i$, a binary indicator $W_i$ of treatment assignment, and a short-term post-treatment outcome $Y_{1i}$. 
The researcher is interested in the effect of the treatment $W_i$ on a long-term post-treatment outcome $Y_{2i}$ that is not measured in the experimental data (i.e., $G_i=E$). 
To complement this deficiency, the researcher obtains an auxiliary observational data set containing measurements for a separate population of individuals $i$ (indicated by $G_i=O$), which consists of the identical list of covariates $X_i$, treatment $W_i$, and short-term outcome $Y_{1i}$, as well as the long-term outcome $Y_{2i}$ that was missing in the experimental data.
The following table summarizes the observability of each of the random variables/vectors in this setup.

    %%%%%%%%%%%%%%%%%%%%%%%%%%%%%%%%%%%%%%%
    \begin{table}[h]\centering
        \renewcommand{\arraystretch}{0.9}
        \begin{tabular}{c|cccc}
            &$Y_{2i}$&$Y_{1i}$&$X_i$&$W_i$\\\hline
            $G_i=E$&$\times$&$\bigcirc$&$\bigcirc$&$\bigcirc$\\\hline
            $G_i=O$&$\bigcirc$&$\bigcirc$&$\bigcirc$&$\bigcirc$
        \end{tabular}
    \end{table}
    %%%%%%%%%%%%%%%%%%%%%%%%%%%%%%%%%%%%%%%

\noindent
The symbol `$\bigcirc$' indicates that the variable is observed by the researcher. 
The symbol `$\times$' indicates that the variable is not observed by the researcher.

Suppose that the latent random vector $(Y_{1i}(0),Y_{1i}(1),Y_{2i}(0),Y_{2i}(1),X_i',W_i,G_i)'$ is randomly drawn from a population distribution, where $Y_{1i}(w)$ (respectively, $Y_{2i}(w)$) denotes the short-term (respectively, long-term) potential outcome under treatment $w \in \{0,1\}$.
The observed outcomes are constructed according to $Y_{1i} = (1-W_i)Y_{1i}(0) + W_iY_{1i}(1)$ for all $i$ and $Y_{2i} = (1-W_i)Y_{2i}(0) + W_iY_{2i}(1)$ for all $i$ such that $G_i=O$.
With these notations, a researcher is interested in identifying some conditional averages $\E[Y_{2i}(1)-Y_{2i}(0)|\mathcal{I}]$ of the long-term treatment effect $Y_{2i}(1)-Y_{2i}(0)$, given some information set $\mathcal{I}$ describing a subpopulation of interest.
Throughout, we maintain the usual potential outcome assumptions such as SUTVA \citep[cf.][]{imbens2015causal}. 

We now introduce the assumptions commonly invoked in the literature of long-term policy evaluation based on data combination.
For simplicity of notations, the subscript $i$ will be omitted hereafter except when it becomes necessary.

%%%%%%%%%%%%%%%%%%%%%%%%%%%%%%%%%%%%%%%
\begin{assumption}[Experimental Internal Validity]\label{ass:IV}
    ${}$\\${}$\hspace{1cm}
    $W \indep Y_2(w), Y_1(w) | X, G=E$ for all $w \in \{0,1\}$.
\end{assumption}
%%%%%%%%%%%%%%%%%%%%%%%%%%%%%%%%%%%%%%%

\noindent
This assumption is plausibly satisfied by construction in cases where the experimental data are collected from a randomized control trial (RCT).
The next assumption, on the other hand, is less commonly imposed and might be questionable in certain applications.

%%%%%%%%%%%%%%%%%%%%%%%%%%%%%%%%%%%%%%%
\begin{assumption}[External Validity of the Experiment]\label{ass:EV}
    ${}$\\${}$\hspace{1cm}
    $G \indep Y_2(w), Y_1(w) | X$ for all $w \in \{0,1\}$.
\end{assumption}
%%%%%%%%%%%%%%%%%%%%%%%%%%%%%%%%%%%%%%%

\noindent
This external validity assumption requires that the distributions of the potential outcomes be identical between the experimental data $(G=E)$ and the observational data $(G=O)$, given pre-treatment covariates $X$.

Throughout the paper, we assume the overlap condition that $\P(W=1|X,G=O) \in (0,1)$ holds almost surely with respect to the law of $X$ given $G=O$.

%%%%%%%%%%%%%%%%%%%%%%%%%%%%%%%%%%%%%%%
\subsection{The Two Alternative Assumptions to Correct for Selection in Observational Data and Their Identifying Formulae}\label{sec:alternative}
%%%%%%%%%%%%%%%%%%%%%%%%%%%%%%%%%%%%%%%

Section \ref{sec:baseline} introduced the basic assumptions concerning the experimental design that are under the control of the researcher.
This section, on the other hand, introduces the key assumptions concerning observational data that are generally not under the control of the researcher.

There are two alternative assumptions for point identification of the long-term average treatment effects in our data setting without additional data requirements -- see Footnotes \ref{foot:additional} and \ref{foot:quantile_quantile}.
They are the latent unconfoundedness (LU) condition \citep{athey2020combining} and the equi-confounding bias (ECB) condition \citep{ghassami2022combining}, as formally stated below.

%%%%%%%%%%%%%%%%%%%%%%%%%%%%%%%%%%%%%%%
\begin{assumption}[Latent Unconfoundedness; \citealp{athey2020combining}] \label{ass:LU}
    ${}$\\${}$\hspace{1cm}
    $W \indep Y_2(w) \mid Y_1(w), X, G=O$ for all $w \in \{0,1\}$.
\end{assumption}
%%%%%%%%%%%%%%%%%%%%%%%%%%%%%%%%%%%%%%%

%%%%%%%%%%%%%%%%%%%%%%%%%%%%%%%%%%%%%%%
\begin{assumption}[Conditional Additive Equi-Confounding Bias; \citealp{ghassami2022combining}] \label{ass:EC}
    ${}$\\${}$\hspace{1cm}
    $\E[Y_2(0)-Y_1(0)|X,W=0,G=O] = \E[Y_2(0)-Y_1(0)|X,W=1,G=O]$.
\end{assumption}
%%%%%%%%%%%%%%%%%%%%%%%%%%%%%%%%%%%%%%%

Observe that Assumption \ref{ass:LU} encompasses the lagged dependent variable (LDV) model as a special case, while Assumption \ref{ass:EC} is essentially the same as the parallel trend condition -- see \citet[][Chapter 5]{angrist2009mostly} for discussions of these two alternative frameworks in the context of program evaluation \textit{without} data combination.
To fix ideas, one can consider, for instance, a simple LDV model:
\begin{align*}
Y_2(0) = \alpha + \rho Y_1(0) + X'\beta + e.
\end{align*}
Assumption \ref{ass:LU} holds for $w=0$ if $e \indep W |Y_1(0), X, G=O$.
On the other hand, Assumption \ref{ass:EC} holds if $\rho=1$ and $e \indep W|X,G=O$.
The former is stronger in one direction, while the latter is stronger in another.
Namely, the former requires conditioning on the LDV for the unconfoundedness, whereas the latter requires $\rho=1$.
Through this illustration, we see that Assumptions \ref{ass:LU} and \ref{ass:EC} are not nested by each other.
%Section \ref{sec:choice} will investigate this relationship from the viewpoints of major treatment selection models.

Under Assumption \ref{ass:LU}, \citet{athey2020combining} nonparametrically identify the long-term average treatment effect $\E[Y_2(1)-Y_2(0)|G=O]$.
More recently, \citet{ghassami2022combining} investigate various long-term treatment effect estimands under each of Assumptions \ref{ass:LU} and \ref{ass:EC}.
Since our objective in this paper is a deeper understanding of the relationships between the alternative assumptions and identifying formulae, instead of exploring a list of various estimands, we focus on a simple estimand, namely the long-term average treatment effect on the treated, defined by
$
\att = \E[Y_2(1)-Y_2(0)|W=1,G=O].
$

Under Assumptions \ref{ass:IV}, \ref{ass:EV}, and \ref{ass:LU}, $\att$ is identified by
\begin{align}
\attlu
=& \E[Y_2|W=1,G=O] + \frac{\P(W=0|G=O) \E[Y_2|W=0,G=O]}{\P(W=1|G=O)}
\notag\\
&- \frac{\E[\E[\E[Y_2|Y_1,W=0,X,G=O]|X,W=0,G=E]|G=O]}{\P(W=1|G=O)}.
\label{eq:attlu}
\end{align}
\citet{athey2020combining} focus on the average treatment effect (ATE), but their proof strategy directly applies to the identification of $\att$ by \eqref{eq:attlu} as well.
%{\color{red}[SENSITIVITY: The main implication of the LU condition (Assumption \eqref{ass:LU}) used to derive \eqref{eq:attlu} is $\E[Y_2(0)|Y_1(0),X,G=O]=\E[Y_2(0)|Y_1(0),W=0,X,G=O]$. Violation of the LU condition by $\left\vert\E[Y_2(0)|Y_1(0),X,G=O]-\E[Y_2(0)|Y_1(0),W=0,X,G=O]\right\vert \leq \Delta$ will translate to $\left\vert\attlu-\att\right\vert \leq \Delta/\P(W=1|G=O).$]}
On the other hand, Under Assumptions \ref{ass:IV}, \ref{ass:EV}, and \ref{ass:EC}, $\att$ is identified by
\begin{align}
\attec
=& \E[Y_2|W=1,G=O]
+ \E\left[\left. \frac{\E[Y_1|X,W=0,G=O]}{\P(W=1|X,G=O)} \right\vert W=1,G=O\right]
\notag\\
&- \E\left[\left. \frac{\E[Y_1|X,W=0,G=E]}{\P(W=1|X,G=O)} + \E[Y_2|X,W=0,G=O] \right\vert W=1, G=O\right].
\label{eq:attec}
\end{align}
See Theorem 5 of \citet{ghassami2022combining}.

An important point to note here is that the estimands, \eqref{eq:attlu} and \eqref{eq:attec}, are different from each other.
If Assumption \ref{ass:LU} holds but Assumption \ref{ass:EC} does not, then $\attlu$ correctly identifies $\att$ but $\attec$ fails to identify $\att$ in general.
In contrast, if Assumption \ref{ass:EC} holds but Assumption \ref{ass:LU} does not, then $\attec$ correctly identifies $\att$ but $\attlu$ fails to identify $\att$ in general.
Since Assumptions \ref{ass:LU} and \ref{ass:EC} do not nest each other as argued above, there does not seem to exist a dominant strategy for a researcher as to which assumption and identifying formula are more robust to employ in practice.

This observation raises a couple of questions.
First, given that a researcher may not know which of the alternative assumptions is more plausible for a given application, is there a way to bound the true $\att$ by utilizing \eqref{eq:attlu} or \eqref{eq:attec}?
We are going to address this question theoretically in Section \ref{sec:bracketing} and empirically in Section \ref{sec:application:bracketing}.
Second, when a point estimate is required, how can we systematically accumulate credible  and generalizable insights, guided by the exogenous variation of experiments with past empirical economic evidence, to make the right choice between \eqref{eq:attlu} and \eqref{eq:attec} in a given application? %, inspired by \citet{lalonde1986evaluating} and sensitivity analysis?
This question will be addressed via \citeauthor{lalonde1986evaluating}-style exercises in Section \ref{sec:application:lalonde}, and through the lens of a nonparametric selection model and sensitivity analysis in Sections \ref{sec:choice} and \ref{sec:sensitivity}.

%%%%%%%%%%%%%%%%%%%%%%%%%%%%%%%%%%%%%%%
\section{The Main Result: Bracketing}\label{sec:bracketing}
%%%%%%%%%%%%%%%%%%%%%%%%%%%%%%%%%%%%%%%

As emphasized in Section \ref{sec:setup}, neither of Assumptions \ref{ass:LU} and \ref{ass:EC} nests the other. 
These assumptions are not empirically testable.
They are not under the researcher's control either.
The researcher needs to make a decision where there is no dominantly more robust choice.

If the latent unconfoundedness (LU) condition (Assumption \ref{ass:LU}) does not hold, then the estimated $\attlu$ in \eqref{eq:attlu} fails to identify the true $\att$.
If the equi-confounding bias (ECB) condition (Assumption \ref{ass:EC}) does not hold, then the estimated $\attec$ in \eqref{eq:attlu} fails to identify the true $\att$.
Hence, in the absence of knowledge about the true data-generating process, there is some risk of bias associated with each of the two estimands.
Therefore, we establish a bracketing relationship between the two estimands, $\attlu$ and $\attec$, analogous to \citet{angrist2009mostly} and \citet{ding2019bracketing}.
While \citet{angrist2009mostly} or \citet{ding2019bracketing} do not consider data combination, our result is tailored to the new data combination setting introduced by \citet{athey2020combining}.

We suppress pre-treatment covariates $X$ for clarity of exposition, as they do not play a role in the identification.
Let us introduce the short-hand auxiliary notation 
$$
\Psi(y) = \E[Y_2(0) - Y_1(0) | Y_1(0)=y,W=0,G=O].
$$
With this notation, we state two assumptions.

%%%%%%%%%%%%%%%%%%%%%%%%%%%%%%%%%%%%%%%
\begin{assumption}[Non-Explosive Process]\label{ass:psi}
$\Psi$ is a non-increasing function.
\end{assumption}
%%%%%%%%%%%%%%%%%%%%%%%%%%%%%%%%%%%%%%%

This assumption requires that the series $\{Y_t(0)\}_t$ of the potential outcomes be non-explosive.
As one particular example, we can interpret Assumption \ref{ass:psi} in the context of the aforementioned lagged dependent variable (LDV) model \citep[][Ch. 5.3]{angrist2009mostly}.
Specifically, consider the LDV model:
$$
Y_2(0) = \alpha + \rho Y_1(0) + e, \quad \E[e|Y_1(0),W=0,G=O]=0.
$$
Since $\Psi(y) = \alpha - (1-\rho) y$ in this example, Assumption \ref{ass:psi} holds under the non-explosive condition $\rho \leq 1$, i.e., the weak sub-unit root condition.
\citet[][pp. 185]{angrist2009mostly} also make this assumption in deriving their bracketing result.

%%%%%%%%%%%%%%%%%%%%%%%%%%%%%%%%%%%%%%%
\begin{assumption}[First-Order Stochastic Dominance]\label{ass:SD}
One of the following holds.\\
${}$\hspace{1cm}(i) $F_{Y_1(0)|W=0,G=O}(y) \leq F_{Y_1(0)|G=E}(y)$ $y \in \mathbb{R}$; or\\
${}$\hspace{1cm}(ii) $F_{Y_1(0)|W=0,G=O}(y) \geq F_{Y_1(0)|G=E}(y)$ $y \in \mathbb{R}$.
\end{assumption}
%%%%%%%%%%%%%%%%%%%%%%%%%%%%%%%%%%%%%%%

Assumption \ref{ass:SD} is empirically testable, as $F_{Y_1(0)|W=0,G=O}=F_{Y_1|W=0,G=O}$ is true and $F_{Y_1(0)|G=E}=F_{Y_1(0)|W=0,G=E}=F_{Y_1|W=0,G=E}$ is also true by the internal validity condition (Assumption \ref{ass:IV}).
While it is testable from empirical data, we remark that condition (i) is more plausible in general.
Condition (i) requires that the distribution of the short-run potential outcome $Y_1(0)$ with no treatment among those who voluntarily chose not to be treated in the observational data (i.e., $W=0$ and $G=O$) first-order stochastically dominates that among the average individual in the experimental group ($G=E$).
This is a reasonable assumption given that rational agents who opt out from treatment may well tend to have higher potential outcomes $Y_1(0)$ with no treatment.
\citet[][pp. 185]{angrist2009mostly} also make a similar assumption in deriving their bracketing result.\footnote{In our notations, \citet[][pp.185]{angrist2009mostly} assume a non-positive covariance between $W_i$ and $Y_1(0)$. In other words, untreated individuals tend to have higher $Y_1(0)$ than treated individuals. Note that $Y_1=Y_1(0)$ is true for all individuals in their data setting, which is distinct from ours.}
With this said, we once again emphasize the empirical testability of Assumption \ref{ass:SD} (i) and (ii).
Indeed, we check that Assumption \ref{ass:SD} (i) is satisfied in our real data analyses to be presented in Section \ref{sec:application}.

Under these two restrictions, we obtain the following bracketing relationship.

%%%%%%%%%%%%%%%%%%%%%%%%%%%%%%%%%%%%%%%
\begin{theorem}[Bracketing]\label{theorem:bracketing}
Suppose that Assumptions \ref{ass:IV}, \ref{ass:psi} and \ref{ass:SD} are satisfied, and that $\E[|Y_t(0)|]<\infty$ holds for each $t \in \{1,2\}$.
Then,  
\begin{align*}
\attlu \leq \attec
&\text{ 
 holds under Assumption \ref{ass:SD} (i), and 
}
\\
\attlu \geq \attec
&\text{
 holds under Assumption \ref{ass:SD} (ii).
}
\end{align*}
\end{theorem}
%%%%%%%%%%%%%%%%%%%%%%%%%%%%%%%%%%%%%%%

\noindent
A proof is relegated to Appendix \ref{sec:theorem:bracketing}.

Note that we invoke the experimental internal validity condition (Assumption \ref{ass:IV}) in this theorem, in addition to Assumptions \ref{ass:psi} and \ref{ass:SD}.
As remarked in Section \ref{sec:baseline}, Assumption \ref{ass:IV} is plausibly satisfied by construction in cases where the experimental data are collected from a randomized control trial (RCT).

\bigskip
\noindent
{\bf Implications of Theorem \ref{theorem:bracketing}:}
As remarked below the statement of Assumption \ref{ass:SD}, the conditions (i) and (ii) are empirically testable.
That is, the researcher knows the direction of the inequality in the bracketing relationship stated in Theorem \ref{theorem:bracketing}.
Suppose that condition (i) is true,\footnote{If condition (ii) is true, the subsequent discussions apply with the directions of the inequalities reversed.} as it is more plausible in general and is also the case with our real-data analysis to be presented in Section \ref{sec:application}.
In this case, we have $\attlu \leq \attec$.
If the LU condition (Assumption \ref{ass:LU}) is true, then $\attlu = \att \leq \attec$ follows.
If the ECB condition (Assumption \ref{ass:EC}) is true, on the other hand, then $\attlu \leq \att = \attec$ follows.
Hence, if either one of the alternative conditions is true, then we enjoy the doubly-robust bound
\begin{align*}
    \attlu \leq \att \leq \attec.
\end{align*}
In other words, $\attlu$ weakly underestimates the true $\att$, while $\attec$ weakly overestimates the true $\att$ regardless of which of the two alternative conditions holds.
This doubly-robust bracketing relation implies that both the LU-based and ECB-based approaches are useful, and should not be considered as mutually exclusive alternatives in practice.
Nonetheless, researchers often want a point estimand rather than bounds.
Since underestimation is a more conservative direction of bias than overestimation, in the sense that scientists normally put a higher priority on Type I Error than Type II Error, it is more robust to assume the LU condition (Assumption \ref{ass:LU}) and use $\attlu$ when non-negative treatment effects are expected.\footnote{Given the conventional bracketing relation \citep[][Section 5.4]{angrist2009mostly}, empirical researchers often take a stance that the LDV is conservatively more robust than the DID. See \citet{crozet2017should}, \citet{glynn2017front}, and \citet{marsh2022trauma} for instance. Since the LDV-based (respectively, DID-based) estimand serves as the lower (respectively, upper) bound in their context, our statement made in the main text reflects the viewpoints of these empirical researchers.}

\medskip
\noindent
{\bf Contributions of Theorem \ref{theorem:bracketing} to the Literature:}
This bracketing result is analogous to that for the policy evaluation \textit{without} data combination extensively discussed by \citet[][Sections 5.3--5.4]{angrist2009mostly}.
Namely, they show that the fixed-effect (or DID) estimator weakly overestimates the true treatment effect when the LDV assumption is true; and the LDV estimator weakly underestimates the true treatment effect when the parallel trend assumption is true.
While \citet{angrist2009mostly} present this relationship for linear models, a more recent paper by \citet{ding2019bracketing} shows this relationship in the context of nonparametric models.
Our Theorem \ref{theorem:bracketing}, therefore, can be considered as a counterpart of \citet{ding2019bracketing}, where we focus on the novel framework of policy evaluations \textit{with} combined experimental and observational data.
Since this novel setting yields more complicated identifying formulae \citep{athey2020combining,ghassami2022combining} than the conventional setting \textit{without} data combination, the technical value as well as the substantive value added by our new bracketing relation is non-trivial relative to the conventional bracketing relationship.

%%%%%%%%%%%%%%%%%%%%%%%%%%%%%%%%%%%%%%%
\section{Empirical Evidence}\label{sec:application}
%%%%%%%%%%%%%%%%%%%%%%%%%%%%%%%%%%%%%%%
This section demonstrates the bracketing relation (Theorem \ref{theorem:bracketing}) using real data consisting of combined experimental and observational data.
We use the same data sets as \citet{athey2020combining} to this end but provide a brief description of them for completeness in Section \ref{sec:data} below.

%%%%%%%%%%%%%%%%%%%%%%%%%%%%%%%%%%%%%%%
\subsection{Data Description}\label{sec:data}
%%%%%%%%%%%%%%%%%%%%%%%%%%%%%%%%%%%%%%%
\subsubsection{Experimental Data: Project STAR}\label{sec:experimental}
%%%%%%%%%%%%%%%%%%%%%%%%%%%%%%%%%%%%%%%
Project STAR (Student/Teacher Achievement Ratio) was a landmark educational experiment conducted in Tennessee from 1985 to 1989. The primary objective of the experiment was to understand the impact of class size on student achievement, particularly focusing on lower-income schools.

Project STAR was motivated by the need to explore whether smaller class sizes could improve educational outcomes in lower socioeconomic settings. In the initial year (1985--86), 6,323 kindergarten students across 79 schools were randomly assigned to small classes (13-17 students) or regular-sized classes (20-25 students). This random assignment was intended to persist through the third grade. The study faced attrition as students moved or were held back in grades. Moreover, students joining in grades 1--3 were also randomly assigned to classes, making the school-by-entry-grade the primary randomization pool. Both students and teachers were randomly assigned to classes. The study administered the Stanford Achievement Test annually to assess math and reading performance, as the state tests did not extend to early grades. 

%%%%%%%%%%%%%%%%%%%%%%%%%%%%%%%%%%%%%%%
\subsubsection{Observational Data: Large-scale Administrative Data}\label{sec:observational}
%%%%%%%%%%%%%%%%%%%%%%%%%%%%%%%%%%%%%%%
Our observational data are derived from the administrative records of a large urban district. This data set includes information on approximately two million children in grades 3--8, encompassing those born between 1966 and 2001.

The data set comprises around 15 million test scores in English language, arts, and math. Due to changes in the testing regime over the 20 years, including a shift from district-specific to statewide tests and variations in test timing, we have normalized test scores to have a mean of zero and a standard deviation of one by year and grade in line with prior research practices \citep[e.g.,][]{staiger2010searching}. This normalization allows for comparability with other samples nationwide.

%%%%%%%%%%%%%%%%%%%%%%%%%%%%%%%%%%%%%%%
\subsubsection{Key Variables}
%%%%%%%%%%%%%%%%%%%%%%%%%%%%%%%%%%%%%%%
In our analysis, we focus on the following variables common between the experimental and observational data.
First, the long-term outcome \textbf{$Y_2$} represents students' test scores in the long term, specifically standardized test scores (averaging between mathematics and English) at the eighth grade. 
Second, the short-term outcome \textbf{$Y_1$} represents students' test scores in the short term, specifically, the third-grade analog of $Y_2$. 
Third, \textbf{$W$} is a binary indicator of treatment in the form of an assignment to a small class size.
Besides, we use gender, race, and eligibility for free lunch to define subpopulations, where the internal and external validities are assumed within each subpopulation.

%%%%%%%%%%%%%%%%%%%%%%%%%%%%%%%%%%%%%%%
\subsection{Empirical Evidence of the Bracketing Relationship}\label{sec:application:bracketing}
%%%%%%%%%%%%%%%%%%%%%%%%%%%%%%%%%%%%%%%

\begin{figure}
\centering
\begin{tabular}{ll}
(1) All Students & (2) Black Students\\ 
\includegraphics[width=0.475\textwidth]{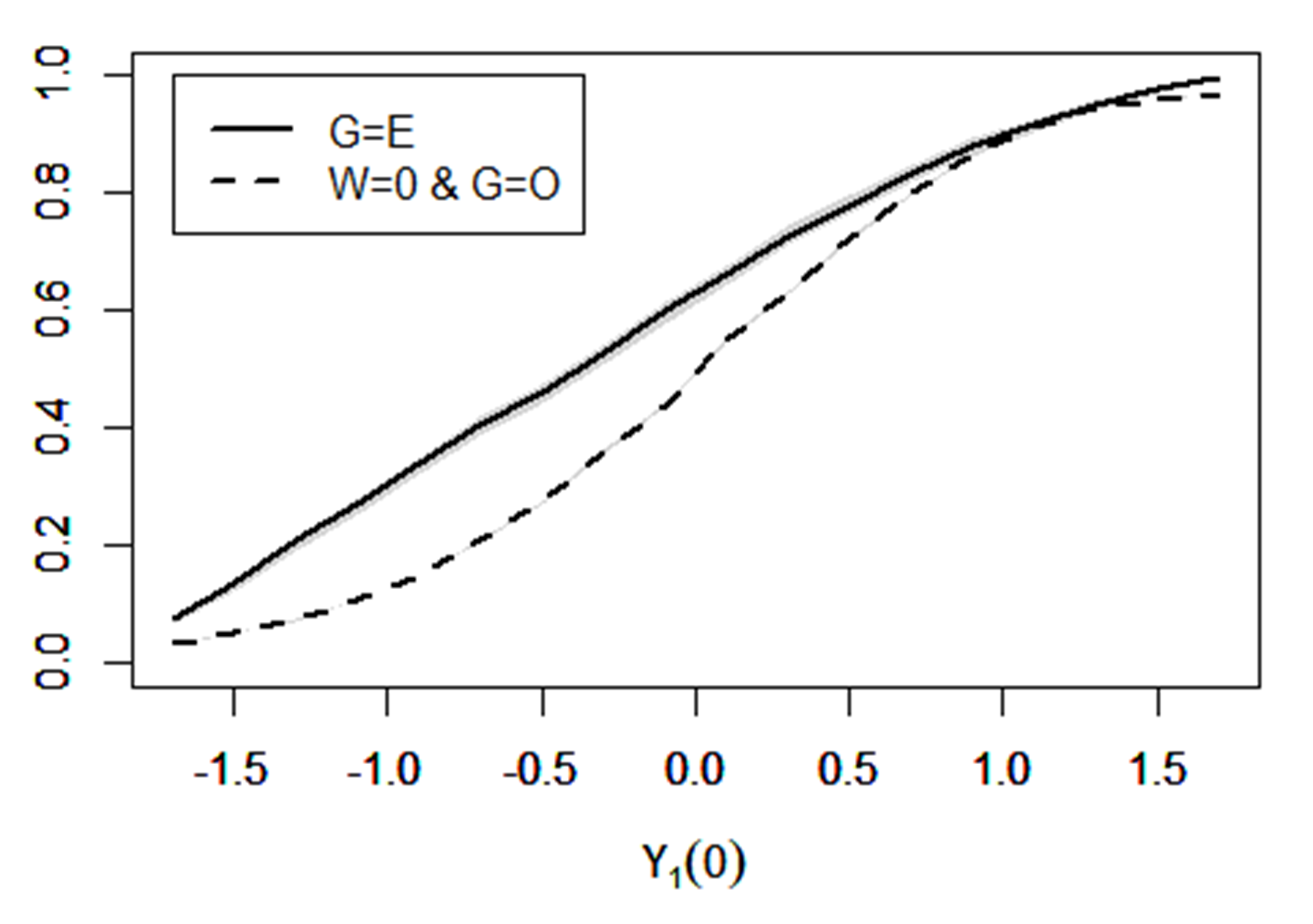}&
\includegraphics[width=0.475\textwidth]{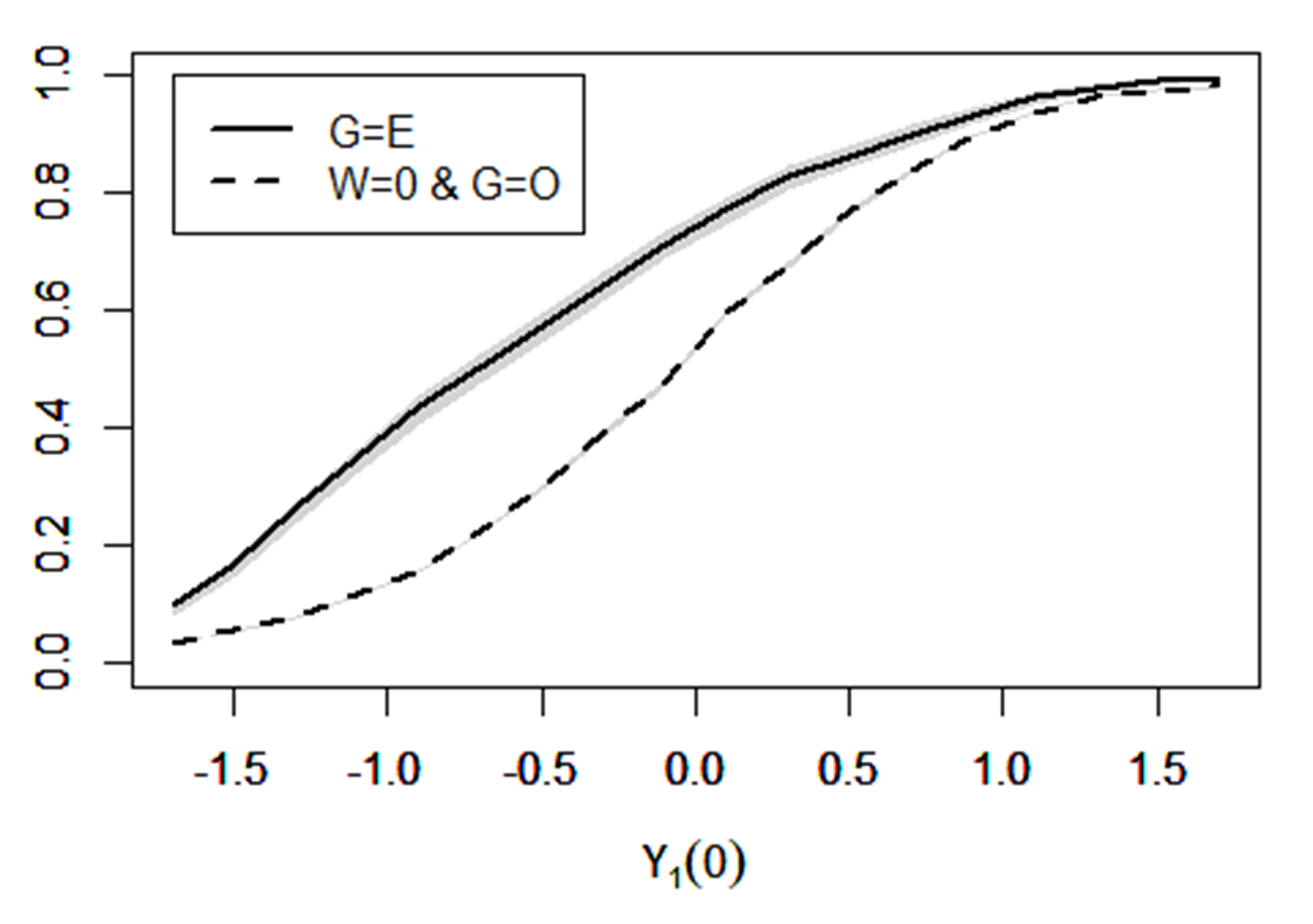}\\\\
(3) Black Male Students & (4) Black Female Students\\ 
\includegraphics[width=0.475\textwidth]{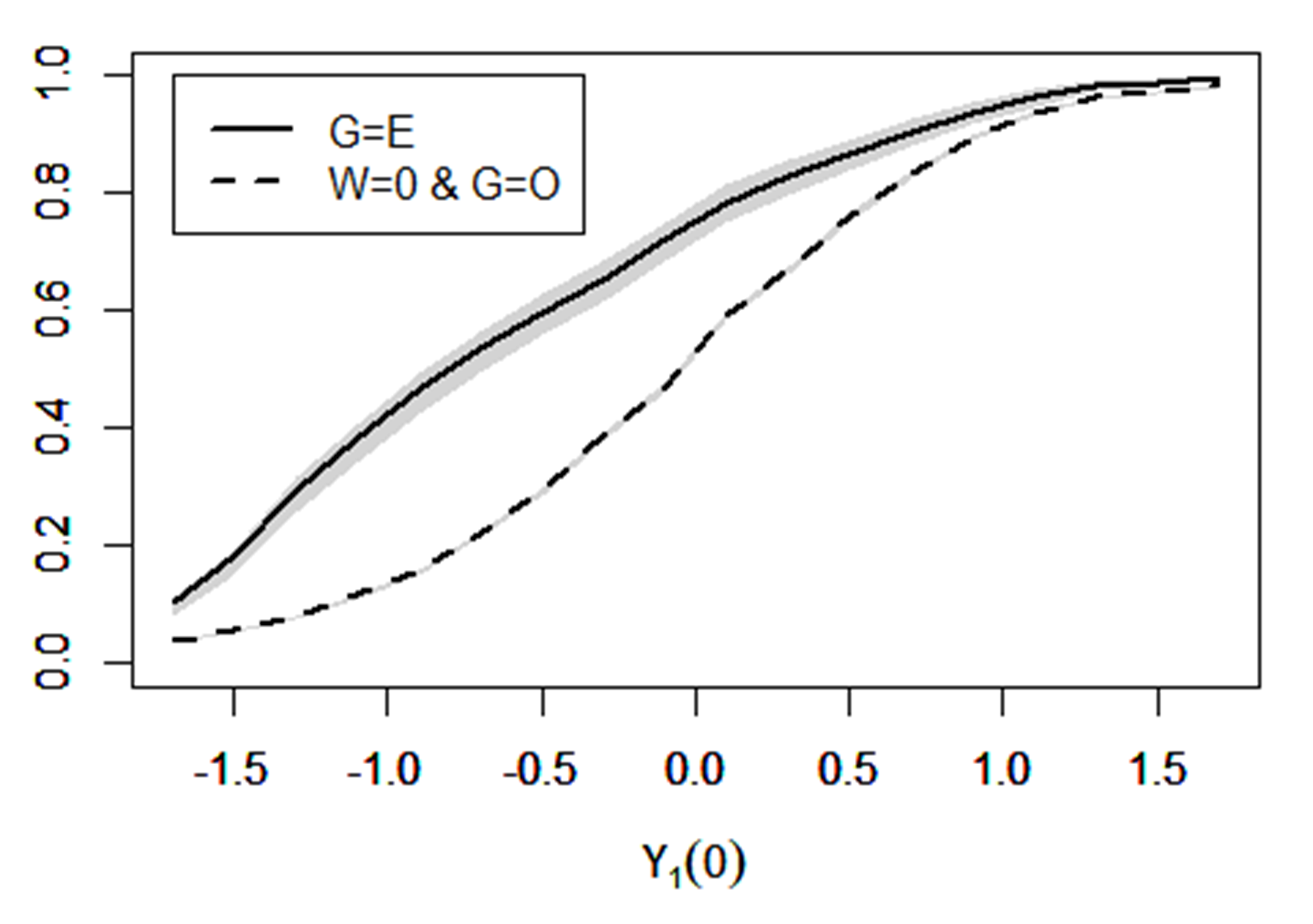}&
\includegraphics[width=0.475\textwidth]{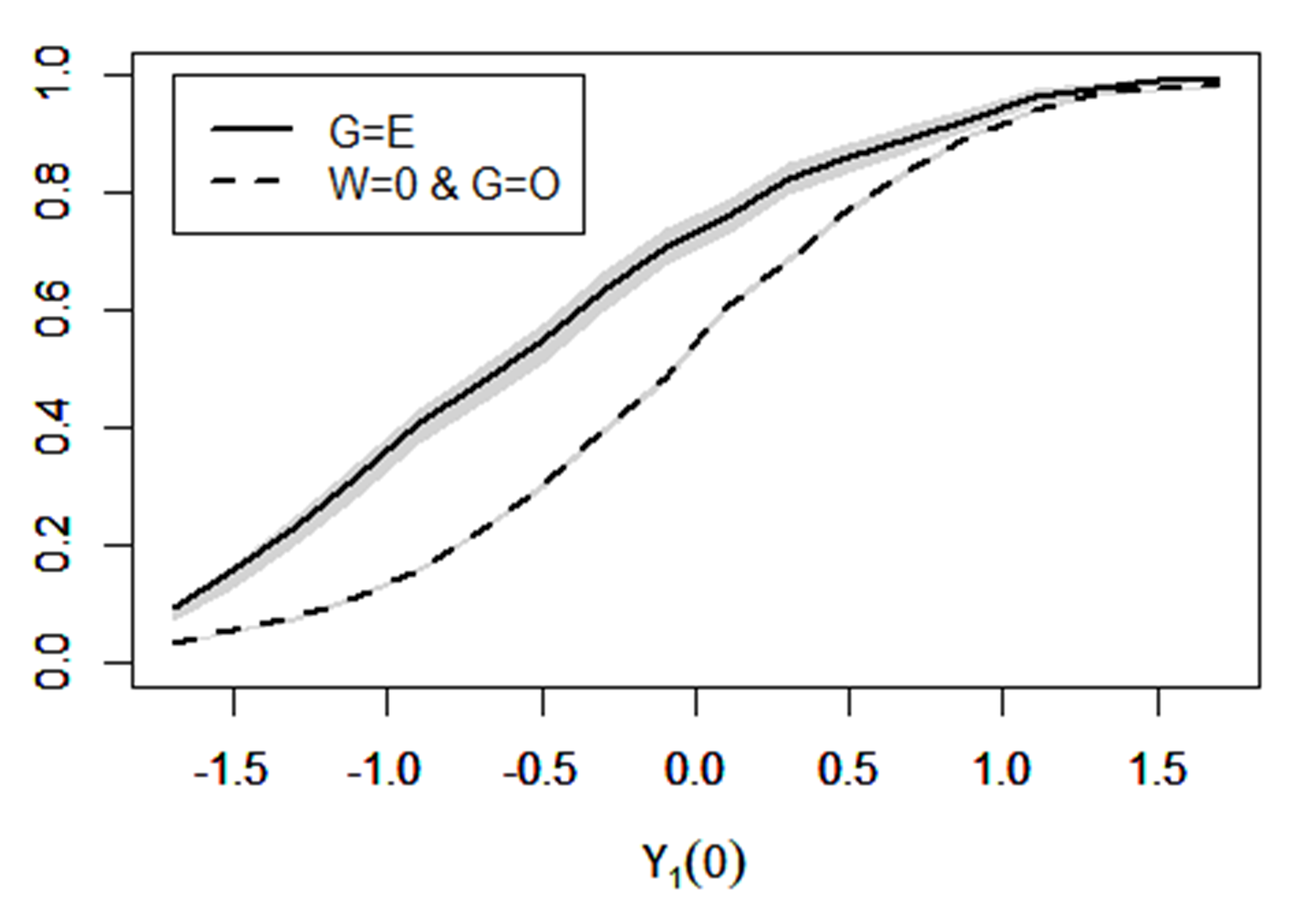}\\
%(5) White Male with Free Lunch & (6) White Female with Free Lunch\\ 
%\includegraphics[width=0.475\textwidth]{fig_X1_0_X2_0_X3_1.png}&
%\includegraphics[width=0.475\textwidth]{fig_X1_1_X2_0_X3_1.png}\\
\end{tabular}
\caption{Graphs of the empirical cumulative distribution functions for
$F_{Y_1(0)|G=E}=F_{Y_1(0)|W=0,G=E}=F_{Y_1|W=0,G=E}$ (solid) and
$F_{Y_1(0)|W=0,G=O}=F_{Y_1|W=0,G=O}$ (dashed) for each of the four subpopulations (1)--(4). The shades indicate point-wise 95\% confidence intervals.}${}$
\label{fig:cdf}
\end{figure}

We are now going to examine our proposed bracketing relationship, as stated in Theorem \ref{theorem:bracketing}.
While Assumption \ref{ass:SD} (i) is plausible as argued in Section \ref{sec:bracketing},
we can empirically check if the stochastic dominance condition is actually satisfied for an application of interest.
Figure \ref{fig:cdf} illustrates the graphs of the empirical cumulative distribution functions for
$F_{Y_1(0)|G=E}=F_{Y_1(0)|W=0,G=E}=F_{Y_1|W=0,G=E}$ (solid) and
$F_{Y_1(0)|W=0,G=O}=F_{Y_1|W=0,G=O}$ (dashed).
Within the target subpopulation of students with lower socio-economic status proxied by eligibility for free lunch, each of the four panels (1)--(4) focuses on a subpopulation characterized by gender and race.\footnote{We show these multiple results, instead of a single aggregated result, to showcase multiple numbers of empirical evidence as opposed to just one.}

Observe that $F_{Y_1(0)|W=0,G=O}(y) \leq F_{Y_1(0)|G=E}(y)$ is likely true for each of the four subpopulations, and Assumption \ref{ass:SD} (i) is plausibly satisfied.
Thus, our main result, Theorem \ref{theorem:bracketing}, implies that the bracketing relation $\attlu \leq \attec$ should hold.
To see if this theoretical prediction is true, we now compute estimates of $\attlu$ and $\attec$.

As we compute $\att$ conditional on each of the subpopulations (1)--(4) characterized by gender and race, we continue to suppress $X$ in our notations.
Thus, \eqref{eq:attlu} and \eqref{eq:attec} can be estimated by
\begin{align*}
\widehat\attlu =& \widehat\E[Y_2|W=1,G=O]
\\
+& \frac{\widehat\P(W=0|G=O) \widehat\E[Y_2|W=0,G=O]}{\widehat\P(W=1|G=O)}
- \frac{\widehat\E [ \widehat\E[Y_2|Y_1,W=0,G=O] |W=0,G=E] }{\widehat\P(W=1|G=O)}
\end{align*}
and
\begin{align*}
\widehat\attec =& \widehat\E[Y_2|W=1,G=O]
\\
+& \frac{ \widehat\E[Y_1|W=0,G=O] }{\widehat\P(W=1|G=O)}
- \frac{ \widehat\E[Y_1|W=0,G=E] }{\widehat\P(W=1|G=O)}
- \widehat\E[Y_2|W=0,G=O],
\end{align*}
respectively.
We can straightforwardly compute $\widehat\E[Y_t|W=w,G=g]$ by the sample conditional mean of $Y_t$ given $W=w$ and $G=g$ for each $t \in \{1,2\}$, $w \in \{0,1\}$, and $g \in \{E,O\}$.
Similarly, we can also straightforwardly compute $\widehat\P(W=w|G=g)$ by the sample conditional probability of $W=w$ given $G=g$ for each $w \in \{0,1\}$ and $g \in \{E,O\}$.
Finally, we compute $\widehat\E[Y_2|Y_1,W=0,G=O]$ by the linear regression of $Y_2$ on $Y_1$ using the subsample of $G=O$ with $W=0$.
Plugging in yields the estimates $\widehat\attlu$ and $\widehat\attec$ reported in the middle two columns of Table \ref{tab:application}.

\begin{table}
\centering
\begin{tabular}{lllccccccccc}
\hline\hline
&&& Na\"ive && \multicolumn{3}{c}{Bracketing} && Experimental\vspace{-1mm}\\
\cline{6-8}
&&& Estimate && LU && ECB && Estimate\vspace{-1mm}\\
&\multicolumn{1}{c}{Subpopulation} & \ \ \ & $\widehat\naive$ & \ \ \ & $\widehat\attlu$ && $\widehat\attec$ & \ & $\widehat\atte$\\
\hline
(1)  & All Students          &&$-0.163$ && 0.186 &$\leq$& 0.479 && 0.181\vspace{-3mm}\\
     &                       &&\ \ $(0.008)$&&(0.022)&      &(0.037)&&(0.044)\\
(2)  & Black Students        &&$-0.148$ && 0.253 &$\leq$& 0.703 && 0.201\vspace{-3mm}\\
     &                       &&\ \ $(0.012)$&&(0.026)&      &(0.046)&&(0.059)\\   
(3)  & Black Male Students   &&$-0.130$ && 0.281 &$\leq$& 0.835 && 0.193\vspace{-3mm}\\
     &                       &&\ \ $(0.017)$&&(0.034)&      &(0.062)&&(0.085)\\
(4)  & Black Female Students &&$-0.163$ && 0.214 &$\leq$& 0.567 && 0.207\vspace{-3mm}\\
     &                       &&\ \ $(0.016)$&&(0.037)&      &(0.064)&&(0.080)\\
\hline
\end{tabular}
\caption{Estimates of $\attlu$, $\attec$, and $\atte$ for each of the four subpopulations (1)--(4). See Section \ref{sec:application:lalonde} for $\atte$. Standard errors are indicated in parentheses.}${}$
\label{tab:application}
\end{table}

Observe that both of the estimates $\widehat\attlu$ and $\widehat\attec$ are positive for each of the subpopulations (1)--(4).
Furthermore, we have $\widehat\attlu \leq \widehat\attec$ for each (1)--(4).
We thus obtained an empirical confirmation of Theorem \ref{theorem:bracketing} with the direction of inequality consistent with the direction of the stochastic dominance (i.e., Assumption \ref{ass:SD} (i)) illustrated in Figure \ref{fig:cdf}.

Moreover, the bracketing relation is informative about the sign and magnitude of causal effects.
To facilitate this discussion, consider the na\"ive observational estimates 
$$
\widehat\naive = \widehat\E[Y_2|W=1,G=O] - \widehat\E[Y_2|W=0,G=O],
$$ 
where $\widehat\E[Y_2|W=w,G=O]$ is the conditional sample mean of $Y_2$ given $W=w$ and $G=O$ for each $w \in \{0,1\}$,
displayed in the first column of Table \ref{tab:application}.
They are significantly negative for all of the four subpopulations.
These results are to be contrasted with the significantly positive estimates by both $\widehat\attlu$ and $\widehat\attec$.
Since the true $\att$ falls between $\attlu$ and $\attec$ as far as either one of the LU and ECB conditions is true, one key takeaway from the bracketing relation provides a credible implication on the sign of the causal effect.
While the lower bound $\widehat\attlu$ ensures the sign as such, the upper bound $\widehat\attec$ additionally informs that the magnitude of the treatment effect is at most a half standard deviation for row (1).

%%%%%%%%%%%%%%%%%%%%%%%%%%%%%%%%%%%%%%%
\subsection{\citeauthor{lalonde1986evaluating}-Style Exercise}\label{sec:application:lalonde}
%%%%%%%%%%%%%%%%%%%%%%%%%%%%%%%%%%%%%%%

While Section \ref{sec:application:bracketing} successfully evidences the bracketing relation and its informativeness about the true causal effects, the large discrepancy between $\widehat\attlu$ and $\widehat\attec$ could be concerning.
Those results motivate us to investigate which of the two alternative assumptions, the LU condition and the ECB condition, is more plausible for this particular application.

To answer this question, we implement analyses \textit{a la} \citet{lalonde1986evaluating}.
Since our experimental data cover the long-term outcome $Y_2$ as well as the short-term outcome $Y_1$, we can in fact compute estimates of the experimental causal effect
\begin{align*}
    \atte = \E[Y_2(1)-Y_2(0)|W=1,G=E] = \E[Y_2|W=1,G=E]-\E[Y_2|W=0,G=E],
\end{align*}
where the second equality follows from the experimental internal validity (Assumption \ref{ass:IV}).
The last column of Table \ref{tab:application} lists its estimate
\begin{align*}
    \widehat\atte = \widehat\E[Y_2|W=1,G=E]-\widehat\E[Y_2|W=0,G=E]
\end{align*}
for each of the subpopulations (1)--(4), where $\widehat\E[Y_2|W=w,G=E]$ is the conditional sample mean of $Y_2$ given $W=w$ and $G=E$ for each $w \in \{0,1\}$.

\begin{table}
\centering
\begin{tabular}{lllccc}
\hline\hline
&\multicolumn{1}{c}{Subpopulation} & \ \ \ & $H_0: \attlu=\atte$ & \ \ \ & $H_0: \attec=\atte$\\
\hline
(1)  & All Students          && Fail to reject $H_0$ && Reject $H_0$\vspace{-3mm}\\
     &                       && (p-value $=0.915$) && (p-value $=0.000$)\\
(2)  & Black Students        && Fail to reject $H_0$ && Reject $H_0$\vspace{-3mm}\\
     &                       && (p-value $=0.427$) && (p-value $=0.000$)\\   
(3)  & Black Male Students   && Fail to reject $H_0$ && Reject $H_0$\vspace{-3mm}\\
     &                       && (p-value $=0.330$) && (p-value $=0.000$)\\
(4)  & Black Female Students && Fail to reject $H_0$ && Reject $H_0$\vspace{-3mm}\\
     &                       && (p-value $=0.936$) && (p-value $=0.000$)\\
\hline
\end{tabular}
\caption{Two-sided tests of the hypotheses $H_0: \attlu=\atte$ and $H_0: \attec=\atte$ for each of the four subpopulations (1)--(4). P-values are indicated in parentheses.}${}$
\label{tab:application_hypothesis}
\end{table}

Observe that $\widehat\atte$ is much closer to $\widehat\attlu$ than to $\widehat\attec$.
To make this observation formal, we conduct two-sided tests of the hypotheses $H_0: \attlu=\atte$ and $H_0: \attec=\atte$.
Table \ref{tab:application_hypothesis} summarizes the test results.
For each of the four subpopulations, we fail to reject the hypothesis $H_0: \attlu=\atte$ but we do reject the hypothesis $H_0: \attec=\atte$.
Hence, we draw the robust conclusion that $\attlu = \atte \neq \attec$ holds.
Recall that our bracketing theory predicts $\attlu = \att \leq \attec$ under the LU condition (Assumption \ref{ass:LU}) but $\attlu \leq \att = \attec$ under the ECB condition (Assumption \ref{ass:EC}).

For this particular application, therefore, the LU condition may be more plausible than the ECB condition.
In addition to our bracketing relationship implying the conservatively greater robustness of $\attlu$ over $\attec$, %this evidence encourages the use of the LU condition when a researcher has to choose one between the two non-nested alternative conditions.
this evidence encourages the use of the LU condition when a researcher has to choose one between the two non-nested alternative conditions for evaluating the effects of policies on educational intervention. We leave explorations of this \citeauthor{lalonde1986evaluating}-style exercise in other applied contexts (e.g., job training programs) as important future work. 

\begin{comment}
Since our experimental data contain the long-term outcome $Y_2$, we can estimate the long-term average treatment effect $\atee = \E[Y_2(1)-Y_2(0)|G=E]$ using the experimental internal validity (Assumption \ref{ass:IV}).
Furthermore, the external validity of the experiment (Assumption \ref{ass:EV}) guarantees $\atee = \ateo = \E[Y_2(1)-Y_2(0)|G=O]$.
Hence, our Lalonde-style exercise is to examine which of $\atelu$ and $\ateec$ is close to $\atee$, where $\atelu$ and $\ateec$ are provided in \citet{athey2020combining} and \citet{ghanem2022selection}, respectively -- see Appendix {\color{red}XXXX} for details.

\begin{align*}
\atelu = 
\E[\E[Y_2|Y_1,W=1,G=O]|G=E]
-
\E[\E[Y_2|Y_1,W=0,G=O]|G=E]
\end{align*}
\citep{athey2020combining}

\begin{align*}
\ateec =
\E[Y_2|W=1,G=O]
-\E[Y_2|W=0,G=O]
+\E[Y_1|W=1,G=E]
\\
-\E[Y_1|W=0,G=E]
+\E[Y_1|W=0,G=O]
-\E[Y_1|W=1,G=O]
\end{align*}
\citep[][Theorem 7]{ghanem2022selection}
\end{comment}

%%%%%%%%%%%%%%%%%%%%%%%%%%%%%%%%%%%%%%%
\section{Characterizing Alternative Identifying Assumptions under Economic Models}\label{sec:choice}
%%%%%%%%%%%%%%%%%%%%%%%%%%%%%%%%%%%%%%%

The \citeauthor{lalonde1986evaluating}-style exercise performed in Seciton \ref{sec:application:lalonde} evidences $\attlu=\atte \neq \attec$.
Since we make additional assumptions (Assumptions \ref{ass:IV} and \ref{ass:EV}), this observation neither formally validates the LU condition (Assumption \ref{ass:LU}) nor formally refutes the ECB condition (Assumption \ref{ass:EC}).
With this said, it suggests that the LU condition (Assumption \ref{ass:LU}) is perhaps more plausible than the ECB condition (Assumption \ref{ass:EC}) in this particular application.
However, we take a step further to provide qualitative reasoning for these empirical findings by asking what \textit{economic substantives} justify the statistical assumption of the LU and ECB conditions.
Our main result is to provide economically interpretable \textit{necessary and sufficient} conditions for the two assumptions through the lens of a \textit{nonparametric family} of selection mechanisms and are not tied to any particular parametric model of selection. 
For concreteness, however, we begin with two particular seminal selection mechanisms, namely \citeauthor{ashenfelter1985susing}'s (\citeyear{ashenfelter1985susing}) Model in Section \ref{sec:ashenfelter} and \citeauthor{roy1951some}'s (\citeyear{roy1951some}) Model in Section \ref{sec:roy}.

Mathematically, we are going to examine
\begin{align}\label{eq:LU}
W \indep Y_2(0) | Y_1(0), G=O,
\end{align}
which corresponds to the latent unconfoundedness (LU) condition (Assumption \ref{ass:LU}) for $w=0$,\footnote{We focus on $w=0$ for ease of writing, but a similar argument follows for $w=1$ as well.} and
\begin{align}\label{eq:EC}
\E[ Y_2(0)-Y_1(0) | W=0, G=O] = \E[ Y_2(0)-Y_1(0) | W=1, G=O],
\end{align}
which corresponds to the equi-confounding bias (ECB) condition (Assumption \ref{ass:EC}).

%%%%%%%%%%%%%%%%%%%%%%%%%%%%%%%%%%%%%%%
\subsection{Through the Lens of \citeauthor{ashenfelter1985susing}'s (\citeyear{ashenfelter1985susing}) Model}\label{sec:ashenfelter}
%%%%%%%%%%%%%%%%%%%%%%%%%%%%%%%%%%%%%%%

\citet{ashenfelter1985susing} propose 
\begin{align}\label{eq:ashenfelter}
W = \mathbbm{1}\{ Y_1(0) + \beta Y_2(0) \leq c \},
\end{align}
as one possible form of selection mechanism,
where $\beta \in [0,1]$ is a discount factor.

Under the selection model \eqref{eq:ashenfelter}, the LU condition \eqref{eq:LU} is written as
\begin{align*}
\mathbbm{1}\{ Y_1(0) + \beta Y_2(0) \leq c \} \indep Y_2(0)|Y_1(0),G=O.
\end{align*}
We can see that this condition generally fails, with $Y_2(0)$ being the factor of dependence.
In the special case where $\beta=0$, however, it reduces to
\begin{align*}
\mathbbm{1}\{ Y_1(0) \leq c \} \indep Y_2(0)|Y_1(0),G=O,
\end{align*}
and it is always satisfied.
Hence, the LU condition \eqref{eq:LU} is plausible under \citeauthor{ashenfelter1985susing}'s (\citeyear{ashenfelter1985susing}) selection model when individuals are perfectly myopic.

Under the selection model \eqref{eq:ashenfelter}, the ECB condition \eqref{eq:EC} is written as
\begin{align*}
&\E[ Y_2(0)-Y_1(0) | \mathbbm{1}\{ Y_1(0) + \beta Y_2(0) \leq c \}=0, G=O] 
\\
=& 
\E[ Y_2(0)-Y_1(0) | \mathbbm{1}\{ Y_1(0) + \beta Y_2(0) \leq c \}=1, G=O].
\end{align*}
We can see that this condition generally fails, with both $Y_1(0)$ and $Y_2(0)$ being the factors of dependence.
In the special case where $\beta=0$, it reduces to
\begin{align*}
\E[ Y_2(0)-Y_1(0) | \mathbbm{1}\{ Y_1(0) \leq c \}=0, G=O] = \E[ Y_2(0)-Y_1(0) | \mathbbm{1}\{ Y_1(0) \leq c \}=1, G=O]
\end{align*}
and it is satisfied if %$\E[Y_2(0)-Y_1(0)|Y_1(0),G=O]=0$.
$Y_2(0)-Y_1(0) \indep Y_1(0) | G=O$.
Hence, the ECB condition \eqref{eq:EC} is plausible under \citeauthor{ashenfelter1985susing}'s (\citeyear{ashenfelter1985susing}) selection model when $\{Y_t(0)\}_t$ follows a martingale-type condition \textit{and} individuals are perfectly myopic.

%%%%%%%%%%%%%%%%%%%%%%%%%%%%%%%%%%%%%%%
\subsection{Through the Lens of \citeauthor{roy1951some}'s (\citeyear{roy1951some}) Model}\label{sec:roy}
%%%%%%%%%%%%%%%%%%%%%%%%%%%%%%%%%%%%%%%

\citet{roy1951some} proposes the selection mechanism of the form
\begin{align}\label{eq:roy}
W = f(Y_1(1)-Y_1(0), Y_2(1)-Y_2(0)).
\end{align}

Under the selection model \eqref{eq:roy}, the LU condition \eqref{eq:LU} is written as
\begin{align*}
f(Y_1(1)-Y_1(0), Y_2(1)-Y_2(0)) \indep Y_2(0)|Y_1(0),G=O.
\end{align*}
We can see that this condition generally fails with $Y_2(0)$ being the factor of dependence.
Even if we impose the myopia assumption that $f$ is constant in the second argument, we still cannot rule out the dependence between $Y_1(1)$ in the first argument of $f$ and $Y_2(0)$.

Under the selection model \eqref{eq:roy}, the ECB condition \eqref{eq:EC} is written as
\begin{align*}
&\E[ Y_2(0)-Y_1(0) | f(Y_1(1)-Y_1(0), Y_2(1)-Y_2(0))=0, G=O] 
\\
=& 
\E[ Y_2(0)-Y_1(0) | f(Y_1(1)-Y_1(0), Y_2(1)-Y_2(0))=1, G=O]
\end{align*}
We can see that this condition generally fails.
In the special case where the potential outcomes take the two-way fixed-effect model of the form
$$
Y_{t}(w) = \alpha_{0} + \lambda_{0t} + \alpha_{1} \lambda_{1t} + \delta_{t}w + \epsilon_{t},
\qquad
\E[\epsilon_{t} | \delta_{1},\delta_{2},G=O]=0 \text{ for all } t \in \{1,2\}
$$
with random $(\alpha_0,\alpha_1,(\delta_t)_{t=1}^2,\varepsilon_t)$ and constant time effects $(\lambda_{0t},\lambda_{1t})_{t=1}^2$,
it reduces to
\begin{align*}
&\E[ \lambda_{02}-\lambda_{01} + \alpha_{1} (\lambda_{12}-\lambda_{11}) | f(\delta_{1}, \delta_{2})=0, G=O] 
\\
=& 
\E[ \lambda_{02}-\lambda_{01} + \alpha_{1} (\lambda_{12}-\lambda_{11}) | f(\delta_{1}, \delta_{2})=1, G=O].
\end{align*}
% Underlying calculation for this formula are as follows:
%$Y_{i2}(0)-Y_{i1}(0) = \lambda_{01}-\lambda_{02} + \alpha_{1i} (\lambda_{02}-\lambda_{01}) + \epsilon_{i2}-\epsilon_{i1}$
%$Y_{i1}(1)-Y_{i1}(0) = \delta_{i1}$
%$Y_{i2}(1)-Y_{i2}(0) = \delta_{i2}$
This reduced restriction is satisfied if $\lambda_{11}=\lambda_{12}$, that is, the interactive time effects are invariant over time.

%%%%%%%%%%%%%%%%%%%%%%%%%%%%%%%%%%%%%%%
\subsection{In A Class of Selection Models with Impefect Foresight}\label{sec:IF}
%%%%%%%%%%%%%%%%%%%%%%%%%%%%%%%%%%%%%%%

\begin{comment}
Section \ref{sec:ashenfelter} reveals that both of the identifying assumptions, \eqref{eq:LU} and \eqref{eq:EC}, entail myopia under the model of \citet{ashenfelter1985susing}.
We provide a sharp characterization of a class of selection models with imperfect foresight in the current subsection.
Unlike the specific models introduced in Sections \ref{sec:ashenfelter}--\ref{sec:roy}, we now consider a nonparametric class of selection models.
We are going to characterize each of \eqref{eq:LU} and \eqref{eq:EC} via a necessary and sufficient condition in this nonparametric family.
The distinct necessary and sufficient conditions between the two alternative identifying conditions unveil the non-nested relationship between them.
\end{comment}

While the previous two subsections shed some light on the economic contents of the two identifying assumptions, some obvious concerns are misspecification of the selection models (also raised by \citet{lalonde1986evaluating}) and logical laxness by being only sufficient but not necessary. In the current subsection, we will explore \textit{necessary and sufficient} conditions under a \textit{nonparametric family of} selection mechanisms, boosting credibility and sharpness of economic characterization. Specifically, we consider a class of selection models of imperfect foresight, motivated by our finding in Section \ref{sec:ashenfelter} that both of the alternative identifying assumptions relate to myopia. % under the model of \citet{ashenfelter1985susing}. %The distinct necessary and sufficient conditions between the two alternative identifying conditions highlight their non-nested relationship.

Reversing the order, we first discuss the ECB condition \eqref{eq:EC} in Section \ref{sec:EC}, as we can take advantage of some results from the existing literature.
This will be followed by our discussion of the LU condition \eqref{eq:LU} in Section \ref{sec:LU}.

%%%%%%%%%%%%%%%%%%%%%%%%%%%%%%%%%%%%%%%
\subsubsection{The ECB Condition  \eqref{eq:EC} under Impefect Foresight}\label{sec:EC}
%%%%%%%%%%%%%%%%%%%%%%%%%%%%%%%%%%%%%%%

As mentioned in Section \ref{sec:alternative}, the ECB condition (Assumption \ref{ass:EC}) is essentially the same as the parallel trend assumption in the context of conventional policy evaluation \textit{without} data combination.
Hence, the work by \citet{ghanem2022selection}, which studies the parallel trend condition from the perspective of an economic selection model, is useful to characterize the ECB condition \eqref{eq:EC}.

\citet{ghanem2022selection} consider the structure
\begin{align}
Y_{t}(0) =& f_t(\alpha,\varepsilon_{t}),\label{eq:selection:y}\\
W =& g(\alpha,\varepsilon_{1},\varepsilon_{2},\nu,\eta_{1},\eta_{2}),\label{eq:selection:w}
\end{align}
and define 
$
\mathcal{G}_{\text{IF}} = \left\{ g : g(\alpha,\varepsilon_{1},\varepsilon_{2},\nu,\eta_{1},\eta_{2}) \text{ is constant in } (\varepsilon_{2},\eta_{2}) \right\}
$. 
This class $\mathcal{G}_{\text{IF}}$ entails imperfect foresight in that the selection indicator $W$ does not account for long-term idiosyncratic components $(\varepsilon_{2},\eta_{2})$.\footnote{\citet{ghanem2022selection} also consider other variants of selection mechanisms. Essentially, they obtain an impossibility result under a general class of selection functions in that the parallel trend condition implies time-invariant relative levels of the long- and short-term potential outcomes. The class $\mathcal{G}_{\text{IF}}$ we focus on is the most general class they consider among those that produce non-trivial characterizations.}
We focus on this class of selection models with imperfect foresight in light of our discovery from Section \ref{sec:ashenfelter} that both of the identifying assumptions, namely the LU condition \eqref{eq:LU} and the ECB condition \eqref{eq:EC}, entail myopia.

\citet{ghanem2022selection} show necessary conditions for the parallel trend condition to hold for all $g \in \mathcal{G}_{\text{IF}}$.
Let $\dot Y_{t}(0) = Y_{t}(0) - \E[Y_{t}(0)|G=O]$.
In our framework \textit{with} combined experimental and observational data, their result directly translates into the following proposition.

\begin{proposition}[\citealp{ghanem2022selection}, Proposition 3.2]\label{prop:ec}
Suppose that there is a subvector $\nu^1$ of $\nu$ such that $\nu^1 \indep (\alpha,\varepsilon_{1},\varepsilon_{2})|G=O$ and $\P(\nu^1 > c|G=O) \in (0,1)$ for some $c \in \mathbb{R}$ in \eqref{eq:selection:y}--\eqref{eq:selection:w}.
Suppose also that either $\P(\E[\dot Y_{2}(0)|\alpha,\varepsilon_{1},G=O]>\dot Y_{1}(0)|G=O)<1$ or $\P(\E[\dot Y_{2}(0)|\alpha,\varepsilon_{1},G=O]<\dot Y_{1}(0)|G=O)<1$ hold.
Then, Equation \eqref{eq:EC} holds for all $g \in \mathcal{G}_{\text{IF}}$ only if $\dot Y_{1}(0)=\E[\dot Y_{2}(0)|\alpha,G=O]$ a.s.
\end{proposition}

This necessary condition also becomes sufficient under an additional condition \citep[cf.][Proposition 3.4]{ghanem2022selection}.
As discussed by \citet{ghanem2022selection}, it can be interpreted as the martingale condition for $\{\dot Y_{t}(0)\}_t$ given $\alpha$ (and also given $G=O$ in our framework with data combination).

%%%%%%%%%%%%%%%%%%%%%%%%%%%%%%%%%%%%%%%
\subsubsection{The LU Condition \eqref{eq:LU} under Impefect Foresight}\label{sec:LU}
%%%%%%%%%%%%%%%%%%%%%%%%%%%%%%%%%%%%%%%

Unlike the ECB condition \eqref{eq:EC} discussed in Section \ref{sec:EC}, the LU condition \eqref{eq:LU} has not been studied in the context of selection models like \eqref{eq:selection:y}--\eqref{eq:selection:w} to the best of our knowledge.
Thus, we present a characterization of the LU condition \eqref{eq:LU} under \eqref{eq:selection:y}--\eqref{eq:selection:w} in the current subsection.

To use $Y_{1}(0)$ as a control variable in the current subsection, we assume the functional form $f_1(\alpha,\varepsilon_{1})=\tilde f_1(\alpha)$ as in \citet{athey2020combining} after suppressing the pre-treatment covariates.
For technical purposes, we further assume that $\tilde f_1$ is measurable, $\mathcal{F}_2$ consists of all measurable functions $f_2$, and $\mathcal{G}_{\text{IF}}$ consists of all measurable functions $g$ with the same restriction from Section \ref{sec:EC}.
To characterize the LU condition \eqref{eq:LU} under \eqref{eq:selection:y}--\eqref{eq:selection:w}, we make the following generalization of the function invertibility.

\begin{definition}[Invertibility in Probability]\label{def:invertibility}
$\tilde f_1$ is invertible in probability if there exists $a(y) \in \tilde f_1^{-1}(\{y\})$ such that $\P( \alpha = a(y) | Y_{1}(0)=y, G=O) = 1$ for every $y$.
\end{definition}

In short, this generalized concept of invertibility requires $\tilde f_1$ to be invertible except on the domain of zero probability.
This invertibility in the structure \eqref{eq:selection:y} can be interpreted as follows.
The production factor $\alpha$ represents the student's current ability in $t=1$, which monotonically determines the potential outcome $Y_1(0)$ in the short run.
In the long run, on the other hand, this short-run ability $\alpha$ together with additional future factors $\varepsilon_2$, not necessarily a scalar, determines the potential outcome $Y_2(0)$ in a way that is not necessarily monotone.
This sense of invertibility characterizes the latent unconfoundedness condition \eqref{eq:LU}, as formally stated in the following proposition.

\begin{proposition}[Necessary and Sufficient Condition for the LU Condition \eqref{eq:LU}]\label{prop:selection}
Suppose that $(\varepsilon_{1},\nu,\eta_{1}) \indep \varepsilon_{2}|\alpha,G=O$ holds in \eqref{eq:selection:y}--\eqref{eq:selection:w} and that the regular conditional probability $\P(\alpha \in A |Y_1(0)=y,G=O)$ exists for each $y \in \text{support}(Y_1(0))$ and each Borel set $A$.
Then, Equation \eqref{eq:LU} holds for all $g \in \mathcal{G}_{\text{IF}}$ and all $f_2 \in \mathcal{F}_2$ if and only if $\tilde f_1$ is invertible in probability.
\end{proposition}

\noindent
A proof is relegated to Appendix \ref{sec:prop:selection}.

This proposition implies that the invertibility of $\tilde f_1$ in the sense of Definition \ref{def:invertibility} is necessary and sufficient for the LU condition \eqref{eq:LU} to hold.
In particular, it does not require restrictions on the relative levels of the long- and short-term potential outcomes unlike the characterization (Proposition \ref{prop:ec}) presented in the previous subsection for the equi-confounding bias condition \eqref{eq:EC}.
This feature of the latent unconfoundedness condition is analogous to the fact that the change-in-changes method \citep{athey2006identification} does not require a restriction on the relative levels of the potential outcomes,\footnote{Indeed, \citet{athey2020combining} describe the latent unconfoundedness as a condition that relates to the change-in-changes method, as is the case with general control function approaches. We move one step further by showing that a slightly modified version is indeed \textit{necessary and sufficient} for the LU condition under general conditions.} unlike the difference-in-differences methods, in the conventional policy evaluations \textit{without} data combination.
In the novel context of long-term policy evaluation \textit{with} data combination, Proposition \ref{prop:selection} provides a similar but different characterization of the LU condition \eqref{eq:LU}.

Recall that the LU condition uses the short-term \textit{potential} outcome $Y_1(0)$, as opposed to the \textit{observed} short-term outcome $Y_1$ as in the conventional surrogacy condition, to control the confounding factor $\alpha$.
As characterized by Proposition \ref{prop:selection}, the LU condition recovers $\alpha = \tilde f_1^{-1}(Y_1(0))$ from the \textit{potential} outcome $Y_1(0)$ via the invertibility of $\tilde f_1$.
On the other hand, $\alpha$ is hard to recover from the \textit{observed} outcome
$Y_1 = \tilde f_1(\alpha)+g(\alpha,\varepsilon_1,\varepsilon_2,\nu,\eta_1,\eta_2)(Y_1(1)-Y_1(0))$ due to the presence of potentially confounding factors other than $\alpha$, even if $\tilde f_1$ were invertible.
This observation through the lens of the economic selection model explains why the LU condition may well be reasonable while the conventional surrogacy condition is not.

In closing this section, we emphasize that the invertibility condition (Definition \ref{def:invertibility}) does not restrict the unobserved factor $\alpha$ to be a scalar.
If a researcher obtains multi-dimensional $Y_1(0)$ then the invertibility condition accordingly admits multidimensional $\alpha$ to characterize the LU condition \eqref{eq:LU}.
Examples include parental background and children's cognitive and non-cognitive skills.
This idea is analogous to the idea of multi-dimensional surrogates proposed by \citet{athey2019surrogate}.

%%%%%%%%%%%%%%%%%%%%%%%%%%%%%%%%%%%%%%%
\subsection{Summary of the Characterized Assumptions}
%%%%%%%%%%%%%%%%%%%%%%%%%%%%%%%%%%%%%%%

We studied the alternative identifying assumptions through the lenses of two parametric classes and one nonparametric class of selection models.
The following table summarizes the observations that we have made in Sections \ref{sec:alternative}, \ref{sec:ashenfelter} \ref{sec:roy}, and \ref{sec:IF}.

%%%%%%%%%%%%%%%%%%%%%%%%%%%%%%%%%%%%%%%
\begin{table}[h]
\centering
\scalebox{0.9}{
\renewcommand{\arraystretch}{0.75}
\begin{tabular}{|l|rl|rl|}
\hline
& \multicolumn{2}{c|}{Latent Unconfoundedness} & \multicolumn{2}{c|}{Equi-Confounding Bias}\\
& \multicolumn{2}{c|}{(Assumption \ref{ass:LU} or \eqref{eq:LU})} & \multicolumn{2}{c|}{(Assumption \ref{ass:EC} or \eqref{eq:EC})}\\
\cline{2-5}
& \multicolumn{2}{c|}{Encompasses LDV} & \multicolumn{2}{c|}{Analogous to parallel trend} \\
\hline
\citeauthor{ashenfelter1985susing}'s & Satisfied under: & myopia & Satisfied under: & myopia $+$ martingale \\
Model of Selection &&&& difference sequence of\\
(Section \ref{sec:ashenfelter}) &&&& potential outcomes\\
\hline
\citeauthor{roy1951some}'s Model of Selection & \multicolumn{2}{l|}{Generally fails.} & Satisfied under: & Two-way fixed-effect \\
(Section \ref{sec:roy}) &&&& model with invariant \\
&&&& interactive time effects\\
\hline
Class of Selection Models & \multicolumn{2}{l|}{Equivalent to function} & Implies: & martingale difference\\
with Imperfect Foresight & \multicolumn{2}{l|}{invertibility in probability,} && sequence of potential\\
(Section \ref{sec:IF})&\multicolumn{2}{l|}{cf. Definition \ref{def:invertibility}.}&&outcomes\\
&\multicolumn{2}{c|}{(Proposition \ref{prop:selection})}&\multicolumn{2}{c|}{(Proposition \ref{prop:ec})}\\
\hline
\end{tabular}
}
%\caption{A summary of the LU and ECB conditions characterized via selection models.}${}$
\label{tab:summary_selection}
\end{table}
%%%%%%%%%%%%%%%%%%%%%%%%%%%%%%%%%%%%%%%

The LU condition (Assumption \ref{ass:LU} or \eqref{eq:LU}) is weaker than the equi-confounding bias condition (Assumption \ref{ass:EC} or \eqref{eq:EC}) under \citeauthor{ashenfelter1985susing}'s model of selection, while the ECB condition (Assumption \ref{ass:EC} or \eqref{eq:EC}) is weaker than the latent unconfoundedness (Assumption \ref{ass:LU} or \eqref{eq:LU}) under \citeauthor{roy1951some}'s model of selection.
Under the nonparametric class of selection models with imperfect foresight, the two alternative conditions are characterized by a distinct set of necessary (and sufficient) conditions.

Recall that the \citeauthor{lalonde1986evaluating}-style exercise in Seciton \ref{sec:application:lalonde} suggests that the LU condition (Assumption \ref{ass:LU}) is consistent with the data.
In other words, the equivalent invertibility condition (Definition \ref{def:invertibility}) may be plausible.
At first glance, this condition appears strong since it implies that the one-dimensional past test score alone captures all the unobserved factors relevant to selection.
However, this observation is in line with empirical findings about the informativeness of student test scores reported in the existing literature.
For instance, in the context of teacher value-added analysis, \citet{chetty2014measuring1} find that accounting for students' prior test scores provides unbiased forecasts of teachers’ impacts on student achievement. Moreover, using the same Project STAR data set as in our analysis, \citet{chetty2011does} show that past test scores serve as excellent controls to predict children's earnings in adulthood. In this sense, our findings reconfirm those in the empirical economics literature.
It is worth noting that the prior test score $Y_1$ \textit{per se} is not sufficient as a surrogate, as noted by \citet[][Section 5]{athey2020combining}, but its latent value in the form of the potential outcome $Y_1(0)$ as in the LU condition may well be a sufficient control.

On the other hand, the \citeauthor{lalonde1986evaluating}-style exercise in Seciton \ref{sec:application:lalonde}
suggests that the ECB condition (Assumption \ref{ass:EC}) is perhaps implausible. 
Thus, under the class of \citeauthor{ashenfelter1985susing}'s model of selection and, more generally, in the nonparametric class of selection with imperfect foresight, the martingale condition is unlikely to hold.
This conclusion is reasonable for the test score as it is supposed to be sub-martingale.\footnote{Note that a conditionally non-degenerate martingale process implies that the variance monotonically increases with $t$. But this contradicts the restriction that the test scores are bounded.}
To embed this into our real-data application, we finally conduct a selection-based sensitivity analysis in the following section.

%%%%%%%%%%%%%%%%%%%%%%%%%%%%%%%%%%%%%%%
\section{Sensitivity Analysis}\label{sec:sensitivity}
%%%%%%%%%%%%%%%%%%%%%%%%%%%%%%%%%%%%%%%

As discussed at the end of the previous section, the \citeauthor{lalonde1986evaluating}-style exercise in Seciton \ref{sec:application:lalonde} implies that the martingale condition, as the necessary (and sufficient) condition for the ECB condition (Assumption \ref{ass:EC}), is likely to fail.
This motivates the following question.
How much deviation from the martingale condition would allow the ECB-type approach to rationalize the experimental estimate $\widehat\atte$ in the \citeauthor{lalonde1986evaluating}-style exercise in Seciton \ref{sec:application:lalonde}?
The current section investigates this question by adapting the idea of the selection-based sensitivity analysis, proposed by \citet[][Section 4]{ghanem2022selection} in the context of the conventional policy evaluation \textit{without} data combination, to our novel context of the long-term policy evaluation \textit{with} data combination.

%%%%%%%%%%%%%%%%%%%%%%%%%%%%%%%%%%%%%%%
\subsection{Sensitivity of the ECB-Based Estimand $\attec$}\label{sec:sensitivity_ecb}
%%%%%%%%%%%%%%%%%%%%%%%%%%%%%%%%%%%%%%%

Consider the following deviation of $\{\dot Y_t(0)\}_t$ from the martingale process, governed by a super-parameter $\overline\rho$.
\begin{align}\label{eq:deviation}
\text{Deviation from Martingale Process:} \quad
\E[\dot Y_2(0)|\alpha,\varepsilon_1,\nu,\eta_1] = \phi(\dot Y_1(0); \overline\rho)
\end{align}
For instance, we can think of the process specified by
\begin{align}\label{eq:deviation:ar}
\E[\dot Y_2(0)|\alpha,\varepsilon_1,\nu,\eta_1] = \phi(\dot Y_1(0)) := \overline\rho \dot Y_1(0).
\end{align}
In this concrete specification, $\overline\rho=1$ entails the martingale process as a special case, whereas $\overline\rho<1$ implies sub-martingale processes.

Let $\dot Y_1 = Y_1 - \E[Y_1|W=0,G=E]$ for a short hand.
The following proposition shows how the super-parameter $\overline\rho$ translates into the discrepancy between the ECB-based estimand $\attec$ and the true value of $\att$.

\begin{proposition}[Sensitivity of $\attec$]\label{prop:sensitivity}
Suppose that Assumptions \ref{ass:IV} and \ref{ass:EV} hold and $\E[|Y_t(w)|]<\infty$ for each $t \in \{1,2\}$ and $w \in \{0,1\}$. Let $g \in \mathcal{G}_{\text{IF}}$ in \eqref{eq:selection:y}--\eqref{eq:selection:w}.
Under the deviation \eqref{eq:deviation} of $\{\dot Y_t(0)\}_t$ from the martingale process,
$
\attec
= \att + \Delta(\overline\rho)
$
holds where
\begin{align*}
\Delta(\overline\rho) = 
\frac{\E[\phi(\dot Y_1; \overline\rho)-\dot Y_1|W=0,G=E]}{\P(W=1|G=O)}
-
\frac{\E[\phi(\dot Y_1; \overline\rho)-\dot Y_1|W=0,G=O]}{\P(W=1|G=O)}.
%\frac{\E[W(\phi(\dot Y_1(0); \overline\rho)-\dot Y_1(0))|G=O]}{\P(W=1|G=O)}
%-\frac{\E[(1-W)(\phi(\dot Y_1(0); \overline\rho)-\dot Y_1(0))|G=O]}{\P(W=0|G=O)}.
\end{align*}
\end{proposition}

A proof is relegated to Appendix \ref{sec:prop:sensitivity}.
In the concrete specification \eqref{eq:deviation:ar} of $\phi$, the bias $\Delta(\overline\rho)$ takes the simple form
\begin{align*}
\Delta(\overline\rho) = (\overline\rho-1)\left[
\frac{\E[\dot Y_1|W=0,G=E]}{\P(W=1|G=O)}
-\frac{\E[\dot Y_1|W=0,G=O]}{\P(W=0|G=O)}\right].
\end{align*}

\bigskip
\noindent
{\bf Contributions of Proposition \ref{prop:sensitivity} to the Literature: }
This result makes non-trivial contributions to the literature on selection-based sensitivity analysis.
\citet[][Section 4]{ghanem2022selection} study the selection-based sensitivity in the context of the conventional policy evaluation \textit{without} data combination.
Following their framework directly, we would obtain
\begin{align*}
\Delta(\overline\rho) =
\frac{\E[\phi(\dot Y_1(0); \overline\rho)-\dot Y_1(0)|G=O]}{\P(W=1|G=O)}
-
\frac{\E[(1-W)(\phi(\dot Y_1(0); \overline\rho)-\dot Y_1(0))|G=O]}{\P(W=0|G=O)\P(W=1|G=O)}.
\end{align*}
In their framework, $Y_1(0)=Y_1$ is observable for all units as the pre-treatment outcome under no anticipation, and hence observational data alone allow for their sensitivity analysis.
On the other hand, in the novel context of long-term policy evaluation \textit{with} data combination as in \citet{athey2020combining}, even the short-run potential outcome $Y_1(0)$ is observed only for the untreated observations, and hence we cannot directly take advantage of the results from the existing literature.
In view of our proof of Proposition \ref{prop:sensitivity}, the reader can see that the internal and external validity conditions (Assumptions \ref{ass:IV} and \ref{ass:EV}) together elegantly solve this unobservability problem.

%%%%%%%%%%%%%%%%%%%%%%%%%%%%%%%%%%%%%%%
\subsection{Revisiting the Empirical Application}\label{sec:sensitivity_application}
%%%%%%%%%%%%%%%%%%%%%%%%%%%%%%%%%%%%%%%

Proposition \ref{prop:sensitivity} motivates the $\overline\rho$-adjusted ECB-based estimate $\widehat\attec(\overline\rho) = \widehat\attec - \widehat\Delta(\overline\rho)$, where $\widehat\Delta(\overline\rho)$ is the sample-counterpart estimator of $\Delta(\overline\rho)$ by replacing the conditional mean (respectively, probability) with the sample conditional mean (respectively, probability).
With this variant $\widehat\attec(\overline\rho)$ of the ECB-based estimate, we will now revisit the empirical application presented in Section \ref{sec:application}.

\begin{figure}[t]
\centering
\includegraphics[width=0.9\textwidth]{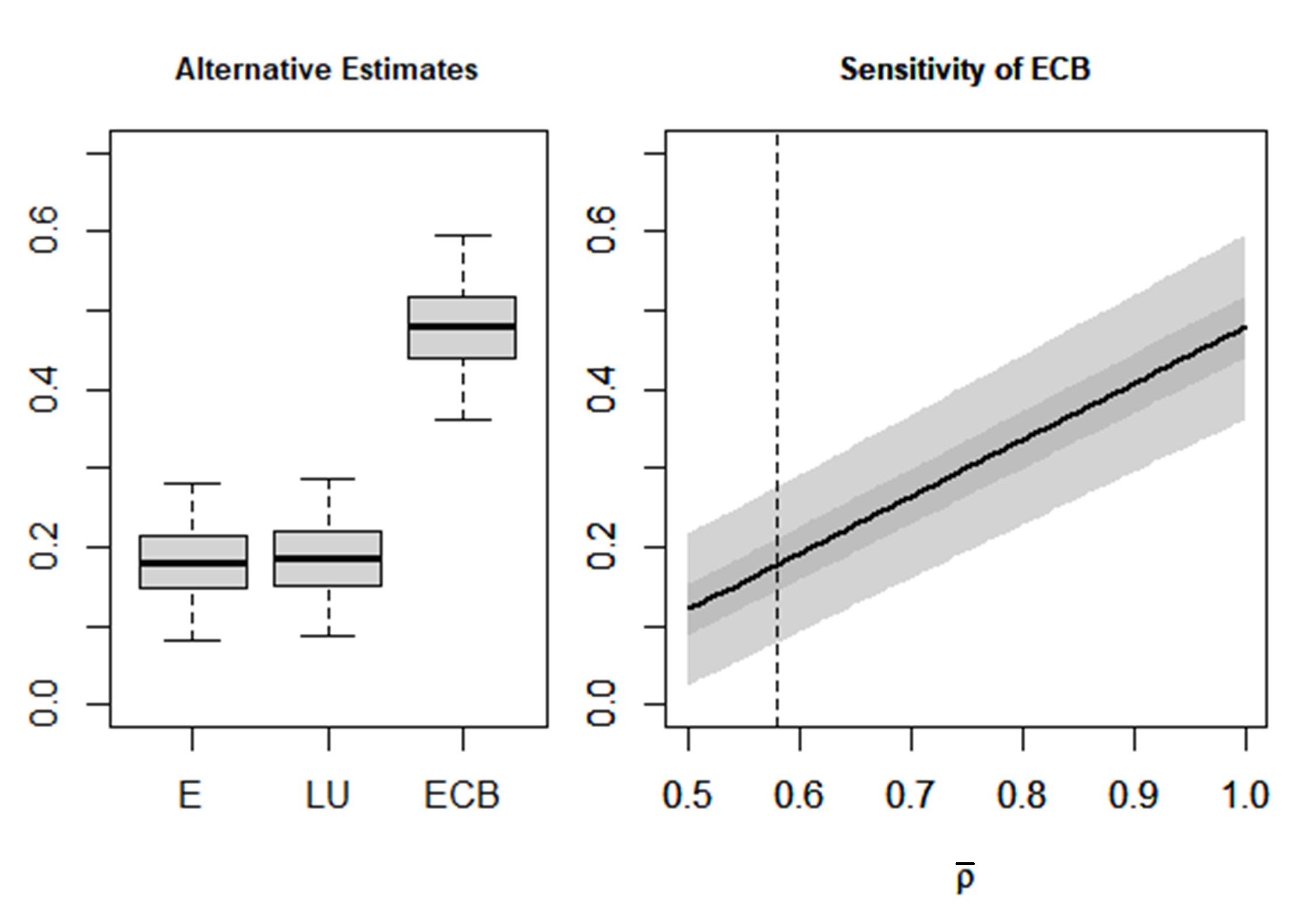}
\caption{Box plots for $(\widehat\atte,\widehat\attlu,\widehat\attec)$ (left) and sensitivity analysis for $\widehat\attec(\overline\rho)$ (right). Displayed around the estimates are the point-wise interquartile ranges and 95\% confidence intervals based on the limit normal approximation. The dashed vertical line on the right panel indicates the value of $\overline\rho$ such that $\widehat\attec(\overline\rho) = \widehat\atte$.}${}$
\label{fig:sensitivity}
\end{figure}

The left panel of Figure \ref{fig:sensitivity} illustrates Table \ref{tab:application} for sample (1) in terms of box plots.
Indicated around the estimates in the middle are the interquartile ranges and 95\% confidence intervals based on the limit normal approximation.
As we observed in Sections \ref{sec:application:bracketing}--\ref{sec:application:lalonde}, the LU-based estimate $\widehat\attlu$ is close to the experimental estimate $\widehat\atte$, but the ECB-based estimate $\widehat\attec$ is far above $\widehat\atte$.

Using the specification \eqref{eq:deviation:ar} for deviations from the martingale condition, the right panel of Figure \ref{fig:sensitivity} illustrates $\widehat\attec(\overline\rho)$ as a function of $\overline\rho \in [0.5,1.0]$.
Recall that $\overline\rho=1$ implies the martingale process, which is consistent with the ECB condition (Assumption \ref{ass:EC}).
Thus, $\widehat\attec(1)$ exactly coincides with the baseline ECB-based estimate $\widehat\attec$.
As $\overline\rho$ becomes smaller, $\widehat\attec(\overline\rho)$ accordingly gets smaller.
It is when $\overline\rho=0.58$ (indicated by the dashed vertical line in the right panel of the figure) that $\widehat\attec(\overline\rho)$ coincides with the experimental estimate $\widehat\atte$.
In other words, for the ECB-type estimate $\widehat\attec(\overline\rho)$ to coincide with the experimental estimate, the martingale condition needs to be violated, and instead be replaced by a sub-martingale condition with the specific value of $\overline\rho=0.58$ in terms of point estimates.
This number is approximately between the white noise process $(\overline\rho-0)$ and the Margingale process $(\overline\rho=1)$.
Thus, the ECB condition, being equivalent to the martingale condition, was perhaps incompatible with this application.

\begin{comment}
\subsection{General Takeaway on the comparative features of the two assumptions}

Write a table here showing that 
LU( essentially invertibility)
the weakness of LU:

requires the confounding factors to be captured to be less than the dimension of the secondary outcome

the strength of LU:

when sufficient-statistics type of variable(e.g., student past test scores) is available, very credible

is scale invariant

ECB( essentially martingale)
the weakness of LU:

scale-variant

martingale may not be suitable more (stationary, or bounded) outcomes of interest

the strenght of LU:

convservatively more robust

can leverage rich empirical literature on martingale hypothesis to see what is reasonable
\end{comment}
%%%%%%%%%%%%%%%%%%%%%%%%%%%%%%%%%%%%%%%
\section{Summary}
%%%%%%%%%%%%%%%%%%%%%%%%%%%%%%%%%%%%%%%
The literature on long-term policy evaluation with data combination proposes alternative identifying assumptions for nonparametric identification of long-term treatment effects.
First, \citet{athey2020combining} propose the latent unconfoundedness (LU) condition.
Second, \citet{ghassami2022combining} propose the equi-confounding bias (ECB) condition, which is analogous to the parallel trend assumption.
The LU and ECB conditions are mutually non-nested and lead to distinct identifying formulae, $\attlu$ and $\attec$, for the true long-term treatment effect $\att$.

In this light, we develop a bracketing relation between the LU-based estimand and the ECB-based estimand.
Specifically, we show that the LU-based estimand $\attlu$ is less than or equal to the ECB-based estimand $\attec$.
Thus, we have $\attlu = \att \leq \attec$
when the LU condition (Assumption \ref{ass:LU}) holds true,
while we have $\attlu \leq \att = \attec$
when the ECB condition (Assumption \ref{ass:EC}) holds true.
More generally, if either the LU condition (Assumption \ref{ass:LU}) \textit{or} the ECB condition (Assumption \ref{ass:EC}) is true, then we enjoy the doubly-robust bound $\attlu \leq \att \leq \attec$.
This result implies that both the LU and ECB approaches are useful, and should not be considered as mutually exclusive alternatives in practice.
If a researcher seeks a point estimand and puts a higher priority on the Type I Error than Type II Error, then
the LU-based estimand $\attlu$ is conservatively more robust than the ECB-based estimand $\attlu$ in the sense that $\attlu$ underestimates $\att$.

The existing literature \citep{angrist2009mostly,ding2019bracketing} provides bracketing relations in the context of policy evaluations \textit{without} data combination.
Our result contributes to this literature by providing a counterpart for long-term policy evaluation with combined experimental and observational data involving more complicated identifying formulae.
%Since the novel setting \textit{with} data combination yields more complicated identifying formulae \citep{athey2020combining,ghassami2022combining} than the conventional setting \textit{without} data combination, the value added by our new bracketing relation is non-trivial relative to the conventional bracketing relationship.

Combining data from Project STAR, a seminal experiment, and extensive administrative observational data, we demonstrate that our proposed bracketing relationship is indeed satisfied.
The implied bounds are informative about the sign and magnitude of the causal effects.
A \citeauthor{lalonde1986evaluating}-style exercise shows that the LU condition may be more plausible than the ECB condition for this particular application of evaluating policies on educational interventions.

We take additional steps to further harness the value of the exogenous variation from the experiment. To understand the economic substantives behind these empirical results, we characterize the LU and ECB conditions from the perspectives of the economic selection model while aiming to mitigate the concern of model misspecification.
Under a general nonparametric class of selection models with imperfect foresight, the LU condition is equivalent to the invertibility of the short-term outcome production function, whereas the ECB condition is equivalent to the martingale process of potential outcomes.

 We thus connect the success of LU condition to the literature on childhood educational interventions \citep{chetty2014measuring1,chetty2011does} documenting the student's past test score as a sufficient statistic \citep{chetty2009sufficient} for latent unobservables.  Crucially, in our new setting under data combination, it is the \textit{latent potential} test score (as in the LU condition), rather than the \textit{observed} test score (as in the conventional surrogacy condition), that plays the role analogous to a sufficient statistic. Likewise, we connect the failure of the ECB condition to the dubious assumption of student test scores evolving as a martingale process.

Finally, anchoring on the hold-out experimental estimate as the baseline truth, we conducted a selection-based sensitivity analysis to find that the martingale condition would need to be violated substantially for the ECB-based estimates to match the true causal effect. 

Table \ref{tab:summary} summarizes the LU and ECB conditions in terms of the bracketing result, its implied robustness, scale sensitivity, the results of our \citeauthor{lalonde1986evaluating}-style exercise, equivalent characterizations, dimension requirements, and recommendations of when to use.

\begin{table}
\renewcommand{\arraystretch}{0.75}
\centering
\begin{tabular}{|l|l|l|}
\cline{2-3}\multicolumn{1}{c|}{}
& Latent Unconfoundedness (LU) & Equi-Confounding Bias (ECB)\\
\hline
Bracketing: & $\attlu = \att \leq \attec$ & $\attlu \leq \att = \attec$\\
\hline
Robustness: & conservatively more robust & vulnerable to overestimation\\
\hline
Scale: & scale-invariant & scale-sensitive\\
\hline
\citeauthor{lalonde1986evaluating}-style & coinciding to the experimental & far above the experimental\\
exercise: & estimates in our application & estimates in our application \\
& $\widehat\attlu \approx \widehat\atte$ & $\widehat\atte \ll \widehat\attec$ \\
\hline
Equivalent to: & function invertibility & martingale property\\
\hline
Dimensionality: & control variables should have & not applicable \\
         & the same (or larger) dimension & \\
\hline
Credible when: & a \textit{latent} sufficient statistic & the potential outcomes follow\\
               & {\small \citep{chetty2009sufficient}} is available & martingale process\\
\hline
\end{tabular}
\caption{A summary of the comparative features of LU and ECB conditions.}${}$
\label{tab:summary}
\end{table}

\begin{comment}
\section{Note}

Note that 
$ECB = ATT + \Delta$,
where $ \Delta = \E[Y_2(0) | W=1,G=O] -\E[Y_2(0)|W=0,G=O] +\frac{ \E[Y_1(0)|W=0,G=O]-\E[Y_1(0)|W=0,G=E]}{\P(W=1|G=O)}$

Note that 
\begin{align*}
&\Delta\\
&=\frac{\E[WY_2(0)|G=O] }{\P(W=1|G=O)} - \frac{\E[(1-W)Y_2(0)|G=O]}{\P(W=0|G=O)} \\&+ \frac{\E[(1-W)Y_1(0)|G=O]}{\P(W=0|G=O)\P(W=1|G=O)} - \frac{\E[Y_1(0)|G=O]}{\P(W=1|G=O)}\\
&=\frac{\E[WY_2(0)|G=O] }{\P(W=1|G=O)} - \frac{\E[(1-W)Y_2(0)|G=O]}{\P(W=0|G=O)} \\&+ \frac{\E[(1-W)Y_1(0)|G=O]}{\P(W=0|G=O)\P(W=1|G=O)} - \frac{\E[WY_1(0)|G=O]}{\P(W=1|G=O)}-\frac{\E[(1-W)Y_1(0)|G=O]}{\P(W=1|G=O)}\\
&=\frac{\E[W(Y_2(0)-Y_1(0)|G=O]}{\P(W=1|G=O)} - \frac{\E[(1-W)Y_2(0)|G=O]}{\P(W=0|G=O)} +\frac{\E[(1-W)Y_1(0)|G=O]}{\P(W=0|G=O)}\\
&= \frac{\E[W(Y_2(0)-Y_1(0)|G=O]}{\P(W=1|G=O)} -\frac{\E[(1-W)(Y_2(0)-Y_1(0)|G=O]}{\P(W=0|G=O)}
\end{align*},
where the first equality uses law of total probability and external validity, the second equality uses$W+(1-W)=1$, the third equality uses the identity $\frac{1}{\P(W=0|G=O)\P(W=1|G=O)}-\frac{1}{\P(W=1|G=O)}=\frac{1}{\P(W=0|G=O)}$

If this is true, we can take the same strategy as \cite{ghanem2022selection}.
\end{comment}

%%%%%%%%%%%%%%%%%%%%%%%%%%%%%%%%%%%%%%%
\vspace{0.5cm}
\appendix
\section*{Appendix}
\section{Mathematical Proofs}
%%%%%%%%%%%%%%%%%%%%%%%%%%%%%%%%%%%%%%%

%%%%%%%%%%%%%%%%%%%%%%%%%%%%%%%%%%%%%%%
\subsection{Proof of Theorem \ref{theorem:bracketing}}\label{sec:theorem:bracketing}
%%%%%%%%%%%%%%%%%%%%%%%%%%%%%%%%%%%%%%%

\begin{proof}
With no pre-treatment covariates, \eqref{eq:attlu} and \eqref{eq:attec} reduce to
\begin{align*}
\attlu =& \E[Y_2|W=1,G=O]
\\
+& \frac{\P(W=0|G=O) \E[Y_2|W=0,G=O]}{\P(W=1|G=O)}
- \frac{\E [ \E[Y_2|Y_1,W=0,G=O] |W=0,G=E] }{\P(W=1|G=O)}
\end{align*}
and
\begin{align*}
\attec =& \E[Y_2|W=1,G=O]
\\
+& \frac{ \E[Y_1|W=0,G=O] }{\P(W=1|G=O)}
- \frac{ \E[Y_1|W=0,G=E] }{\P(W=1|G=O)}
- \E[Y_2|W=0,G=O],
\end{align*}
respectively.
Their difference can be signed by
\begin{align}
&\P(W=1|G=O) \cdot (\attlu - \attec) 
\notag\\
=& 
\E[Y_2|W=0,G=O] - \E[ \E[ Y_2 |Y_1,W=0,G=O] | W=0, G=E]
\notag\\
&- \E[Y_1|W=0,G=O] + \E[Y_1|W=0,G=E]
\notag\\
=& 
\E[Y_2(0) - Y_1(0)|W=0,G=O] - \E[ \E[ Y_2(0) |Y_1,W=0,G=O] - Y_1(0) | W=0, G=E]
\notag\\
=& 
\E[Y_2(0) - Y_1(0)|W=0,G=O] - \E[ \E[ Y_2(0) - Y_1(0) |Y_1(0),W=0,G=O] | W=0, G=E]
\notag\\
=& 
\E[ \Psi(Y_1(0)) |W=0,G=O] - \E[ \Psi(Y_1(0)) | W=0, G=E]
\notag\\
=& 
\E[ \Psi(Y_1(0)) |W=0,G=O] - \E[ \Psi(Y_1(0)) | G=E],
\label{eq:sign}
\end{align}
where 
the first equality uses the law of total probability $\P(W=0|G=O) + \P(W=1|G=O)=1$,
%the second equality uses $\E[Y_1|W=0,G=E]=\E[Y_1(0)|G=E]$ by Assumption \ref{ass:IV},
the third equality uses the equivalence of the sigma-algebras deriving from the logical equivalence $(Y_1\in S,W=0,G=O)\Leftrightarrow (Y_1(0) \in S,W=0,G=O)$ for all Borel sets $S \subset \mathbb{R}$,
the fourth equality uses the law of iterated expectations, and the last equality uses Assumption \ref{ass:IV}.
Under Assumption \ref{ass:psi}, note that Assumption \ref{ass:SD} (i) and Assumption \ref{ass:SD} (ii) imply
\begin{align}
&\E[ \Psi(Y_1(0)) |W=0,G=O] \leq \E[ \Psi(Y_1(0)) | G=E] \qquad\text{and}\label{eq:SDi}
\\
&\E[ \Psi(Y_1(0)) |W=0,G=O] \geq \E[ \Psi(Y_1(0)) | G=E], \label{eq:SDii}
\end{align}
respectively.
Thus, \eqref{eq:sign} and \eqref{eq:SDi} together yield the bracketing relation
$
\attlu \leq \attec
$
under Assumption \ref{ass:SD} (i).
Similarly, \eqref{eq:sign} and \eqref{eq:SDii} together yield the bracketing relation
$
\attlu \geq \attec
$
under Assumption \ref{ass:SD} (ii).
\end{proof}
%%%%%%%%%%%%%%%%%%%%%%%%%%%%%%%%%%%%%%%

%%%%%%%%%%%%%%%%%%%%%%%%%%%%%%%%%%%%%%%
\subsection{Proof of Proposition \ref{prop:selection}}\label{sec:prop:selection}
%%%%%%%%%%%%%%%%%%%%%%%%%%%%%%%%%%%%%%%

\begin{proof}
{\bf ($\Longleftarrow$)}
Suppose that $\tilde f_1$ is invertible in probability.
Then, given $y \in \text{support}(Y_2(0))$, there exists $a(y) \in \tilde f_1^{-1}(\{y\})$ such that $\P( \alpha = a(y) | Y_{1}(0)=y, G=O) = 1$ by Definition \ref{def:invertibility}.
Thus, the structure \eqref{eq:selection:y}--\eqref{eq:selection:w} can be written as
\begin{align}
Y_{2}(0) =& f_2(a(y),\varepsilon_{2}) \qquad\text{and} \label{eq:selection:y_inv}\\
W =& g(a(y),\varepsilon_{1},\varepsilon_{2},\nu,\eta_{1},\eta_{2})\label{eq:selection:w_inv}
\end{align}
almost surely given $Y_{1}(0)=y$ and $G=O$.
If $g_{\text{IF}} \in \mathcal{G}_{\text{IF}}$,
then \eqref{eq:selection:y_inv} and \eqref{eq:selection:w_inv} reduce to
\begin{align}
Y_{2}(0) =& f_{2,\text{IF}}(y,\varepsilon_{2}) := f_2(a(y),\varepsilon_{2})) \qquad\text{and} \label{eq:selection:y_inv_ii}\\
W =& g_{2,\text{IF}}(y,\varepsilon_{1},\nu,\eta_{1}) := g_{\text{IF}}(a(y),\varepsilon_{1},\varepsilon_{2},\nu,\eta_{1},\eta_{2}))\label{eq:selection:w_inv_ii}
\end{align}
almost surely given $Y_{1}(0)=y$ and $G=O$.

Suppose that $\alpha = a(y)$.
Then, by the definition of $a(y) \in \tilde f_1^{-1}(\{y\})$, we have $y = \tilde f_1(\alpha) = Y_1(0)$.
This argument shows the logical implication
\begin{equation}\label{eq:logical_implication}
\alpha=a(y) \Longrightarrow Y_1(0)=y.
\end{equation}

Now, let $A$ and $B$ be any Borel sets with their dimensions the same as those of $(\varepsilon_{1},\nu,\eta_{1})$ and $\varepsilon_{2}$, respectively.
Observe that
\begin{align*}
&\P((\varepsilon_{1},\nu,\eta_{1}) \in A, \varepsilon_{2} \in B | Y_{1}(0)=y,G=O)
\\
=&
\P((\varepsilon_{1},\nu,\eta_{1}) \in A, \varepsilon_{2} \in B | \alpha=a(y),G=O)
\cdot \P(\alpha=a(y) | Y_{1}(0)=y,G=O)
\\
&+\P((\varepsilon_{1},\nu,\eta_{1}) \in A, \varepsilon_{2} \in B | \alpha \neq a(y),G=O)
\cdot \P(\alpha \neq a(y) | Y_{1}(0)=y,G=O)
\\
=& \P((\varepsilon_{1},\nu,\eta_{1}) \in A, \varepsilon_{2} \in B | \alpha=a(y),G=O)
\\
=& \P((\varepsilon_{1},\nu,\eta_{1}) \in A | \alpha=a(y),G=O) \cdot \P(\varepsilon_{2} \in B | \alpha=a(y),G=O)
\\
=& \big[ \P((\varepsilon_{1},\nu,\eta_{1}) \in A | \alpha=a(y),G=O)
\cdot \P(\alpha=a(y) | Y_{1}(0)=y,G=O)
\\
&+\P((\varepsilon_{1},\nu,\eta_{1}) \in A | \alpha \neq a(y),G=O)
\cdot \P(\alpha \neq a(y) | Y_{1}(0)=y,G=O) \big] \cdot
\\
& \big[ \P(\varepsilon_{2} \in B | \alpha=a(y),G=O)
\cdot \P(\alpha=a(y) | Y_{1}(0)=y,G=O)
\\
&+\P(\varepsilon_{2} \in B | \alpha \neq a(y),G=O)
\cdot \P(\alpha \neq a(y) | Y_{1}(0)=y,G=O) \big]
\\
=& \P((\varepsilon_{1},\nu,\eta_{1}) \in A | Y_{1}(0)=y,G=O) \cdot \P(\varepsilon_{2} \in B | Y_{1}(0)=y,G=O)
\end{align*}
holds, where 
the first equality is due to the law of total probability and the logical implication \eqref{eq:logical_implication},
the second equality is due to the invertibility of $\tilde f_1$ in probability,
the third equality follows by the assumption $(\varepsilon_{1},\nu,\eta_{1}) \indep \varepsilon_{2}|\alpha,G=O$ stated in the proposition,
the fourth equality is due to the invertibility of $\tilde f_1$ in probability, and
the last equality is due to the law of total probability.
Thus, it follows that $(\varepsilon_{1},\nu,\eta_{1}) \indep \varepsilon_{2}|Y_{1}(0)=y,G=O$.
With \eqref{eq:selection:y_inv_ii}--\eqref{eq:selection:w_inv_ii}, this conditional independence in turn implies $W \indep Y_{2}(0)|Y_{1}(0)=y,G=O$ by \citet[][Lemma 4.2]{dawid1979conditional}.

\bigskip\noindent
{\bf ($\Longrightarrow$)}
Suppose that $\tilde f_1$ is not invertible in probability.
Then, by Definition \ref{def:invertibility}, there exists a partition $\{A_1,A_2\}$ of the Borel set $\tilde f_1^{-1}(\{y\})$ such that $\P(\alpha \in A_1 | Y_{1}(0)=y,G=O) > 0$ and $\P(\alpha \in A_2 | Y_{1}(0)=y,G=O) > 0$.
Choose $f_2 \in \mathcal{F}_2$ such that $f_2(A_1,\varepsilon_{2})=\{1\}$ and $f_2(A_2,\varepsilon_{2}) = \{0\}$ for all $\varepsilon_{2}$.
Choose $g_{\textit{IF}} \in \mathcal{G}_{\textit{IF}}$ such that $g_{\textit{IF}}(A_1,\varepsilon_{1},\varepsilon_{2},\nu,\eta_{1},\eta_{2}) = \{0\}$ and $g_{\textit{IF}}(A_2,\varepsilon_{1},\varepsilon_{2},\nu,\eta_{1},\eta_{2}) = \{1\}$ for all $(\varepsilon_{1},\varepsilon_{2},\nu,\eta_{1},\eta_{2})$. 
Note that such $g_{\textit{IF}}$ is an element of $\mathcal{G}_{\text{IF}}$ in that it satisfies its restriction.
Observe that we have
\begin{align*}
\E[W Y_{2}(0)|Y_{1}(0)=y,G=O]
=&
\P(\alpha \in A_1|Y_{1}(0)=y, G=O) \cdot \E[W Y_{2}(0)|\alpha \in A_1,G=O]
\\
+&
\P(\alpha \in A_2|Y_{1}(0)=y, G=O) \cdot \E[W Y_{2}(0)|\alpha \in A_2,G=O]
\\
=& 0,
\\
\E[W|Y_{1}(0)=y,G=O]
=&
\P(\alpha \in A_1|Y_{1}(0)=y, G=O) \cdot \E[W|\alpha \in A_1,G=O]
\\
+&
\P(\alpha \in A_2|Y_{1}(0)=y, G=O) \cdot \E[W|\alpha \in A_2,G=O]
\\
=&\P(\alpha \in A_2|Y_{1}(0)=y, G=O) > 0,
\qquad\text{and}
\\
\E[Y_{2}(0)|Y_{1}(0)=y,G=O]
=&
\P(\alpha \in A_1|Y_{1}(0)=y, G=O) \cdot \E[Y_{2}(0)|\alpha \in A_1,G=O]
\\
+&
\P(\alpha \in A_2|Y_{1}(0)=y, G=O) \cdot \E[Y_{2}(0)|\alpha \in A_2,G=O]
\\
=& \P(\alpha \in A_1|Y_{1}(0)=y, G=O) > 0.
\end{align*}
Combining these three sets of equalities together yields
\begin{align*}
\E[W Y_{2}(0)|Y_{1}(0)=y,G=O] = 0 < \E[W|Y_{1}(0)=y,G=O] \cdot \E[Y_{2}(0)|Y_{1}(0)=y,G=O],
\end{align*}
and hence $W \not\indep Y_{2}(0)|Y_{1}(0)=y,G=O$.
\end{proof}

\subsection{Proof of Proposition \ref{prop:sensitivity}}\label{sec:prop:sensitivity}
%%%%%%%%%%%%%%%%%%%%%%%%%%%%%%%%%%%%%%%
\begin{proof}
With no pre-treatment covariates, \eqref{eq:attec} can written as
$
\attec
= \att + \Delta,
$
where
\begin{align*}
\Delta
=& \E[Y_2(0)|W=1,G=O] - \E[Y_2(0)|W=0,G=O] 
\\
&+ \frac{\E[Y_1(0)|W=0,G=O] - \E[Y_1(0)|W=0,G=E]}{\P(W=1|G=O)} 
\end{align*}
Observe that $\Delta$ can be rewritten as
\begin{align}
\Delta
=& \frac{\P(W=1|G=O)\E[WY_2(0)|W=1,G=O]}{\P(W=1|G=O)} 
\notag\\
&- \frac{\P(W=0|G=O)\E[(1-W)Y_2(0)|W=0,G=O]}{\P(W=0|G=O)}
\notag\\
&+ \frac{\P(W=0|G=O)\E[(1-W)Y_1(0)|W=0,G=O]}{\P(W=0|G=O)\P(W=1|G=O)} 
- \frac{\E[Y_1(0)|W=0,G=E]}{\P(W=1|G=O)} 
\notag\\
=& \frac{\E[WY_2(0)|G=O]}{\P(W=1|G=O)} 
- \frac{\E[(1-W)Y_2(0)|G=O]}{\P(W=0|G=O)}
\notag\\
&+ \frac{\E[(1-W)Y_1(0)|G=O]}{\P(W=0|G=O)\P(W=1|G=O)} 
- \frac{\E[Y_1(0)|G=O]}{\P(W=1|G=O)} 
\notag\\
=& \frac{\E[WY_2(0)|G=O]}{\P(W=1|G=O)} 
- \frac{\E[(1-W)Y_2(0)|G=O]}{\P(W=0|G=O)}
\notag\\
&+ \left(\frac{1}{\P(W=0|G=O)} + \frac{1}{\P(W=1|G=O)}\right) \E[(1-W)Y_1(0)|G=O] 
\notag\\
&- \frac{\E[WY_1(0)|G=O] + \E[(1-W)Y_1(0)|G=O]}{\P(W=1|G=O)} 
\notag\\
=& \frac{\E[W(Y_2(0) - Y_1(0))|G=O]}{\P(W=1|G=O)}
-\frac{\E[(1-W)(Y_2(0)-Y_1(0))|G=O]}{\P(W=0|G=O)}
\notag\\
=& \frac{\E[W(\dot Y_2(0) - \dot Y_1(0))|G=O]}{\P(W=1|G=O)}
-\frac{\E[(1-W)(\dot Y_2(0)-\dot Y_1(0))|G=O]}{\P(W=0|G=O)},
\label{eq:delta}
\end{align}
where
the second equality is due to the law of total probability (for the first, second, and third terms) and Assumptions \ref{ass:IV}--\ref{ass:EV} (for the last term),
the third equality uses the property
$\P(W=0|G=O) + \P(W=1|G=O) = 1$ of the probability (for the third term) and the algebraic identity $W+(1-W)=1$,
the fourth equality combines terms with the common denominator, and
the fifth equality is due to a cancellation of $(\E[Y_2(0)-Y_1(0)|G=O])$ between the two terms.

Now, the numerator in the first term on the right-hand side of \eqref{eq:delta} can be rewritten as
\begin{align*}
\E[W(\dot Y_2(0)-\dot Y_1(0))|G=O]
=&
\E[\E[W(\dot Y_2(0)-\dot Y_1(0))|\alpha,\varepsilon_1,\nu,\eta_1]|G=O]
\\
=&
\E[W(\E[\dot Y_2(0)|\alpha,\varepsilon_1,\nu,\eta_1]-\dot Y_1(0))|G=O]
\\
=&
\E[W(\phi(\dot Y_1(0); \overline\rho)-\dot Y_1(0))|G=O],
\end{align*}
where
the first equality is due to the law of iterated expectations,
the second equality follows from $g \in \mathcal{G}_{\text{IF}}$ and equations \eqref{eq:selection:y}--\eqref{eq:selection:w}, and
the third equality is due to 
\eqref{eq:deviation}.
Similarly, the numerator in the second term on the right-hand side of \eqref{eq:delta} can be rewritten as
\begin{align*}
\E[(1-W)(\dot Y_2(0)-\dot Y_1(0))|G=O]
=
\E[(1-W)(\phi(\dot Y_1(0); \overline\rho)-\dot Y_1(0))|G=O].
\end{align*}
Therefore, $\Delta$ displayed in \eqref{eq:delta} takes the form of
\begin{align*}
\Delta(\overline\rho) 
=& 
\frac{\E[W(\phi(\dot Y_1(0); \overline\rho)-\dot Y_1(0))|G=O]}{\P(W=1|G=O)}
-\frac{\E[(1-W)(\phi(\dot Y_1(0); \overline\rho)-\dot Y_1(0))|G=O]}{\P(W=0|G=O)}
\\
=& 
\frac{\E[\phi(\dot Y_1(0); \overline\rho)-\dot Y_1(0)|G=O]}{\P(W=1|G=O)}
-
\frac{\E[(1-W)(\phi(\dot Y_1(0); \overline\rho)-\dot Y_1(0))|G=O]}{\P(W=0|G=O)\P(W=1|G=O)}
\\
=& 
\frac{\E[\phi(\dot Y_1(0); \overline\rho)-\dot Y_1(0)|W=0,G=E]}{\P(W=1|G=O)}
-
\frac{\E[\phi(\dot Y_1(0); \overline\rho)-\dot Y_1(0)|W=0,G=O]}{\P(W=1|G=O)}
\\
=& 
\frac{\E[\phi(\dot Y_1; \overline\rho)-\dot Y_1|W=0,G=E]}{\P(W=1|G=O)}
-
\frac{\E[\phi(\dot Y_1; \overline\rho)-\dot Y_1|W=0,G=O]}{\P(W=1|G=O)}
\end{align*}
where 
the second equality uses the property
$\P(W=0|G=O) + \P(W=1|G=O) = 1$ of the probability,
the third equality is due to Assumptions \ref{ass:IV} and \ref{ass:EV} (for the first term) and the law of total probability (for the second term), and the last equality uses $\E[Y_1(0)|G=O]=\E[Y_1(0)|G=E]=\E[Y_1|W=0,G=E]$ by Assumption \ref{ass:IV} and \ref{ass:EV}.
\end{proof}

%%%%%%%%%%%%%%%%%%%%%%%%%%%%%%%%%%%%%%%
\setlength{\baselineskip}{6.8mm}
\bibliographystyle{apalike}
\bibliography{reference}

\end{document}